\documentclass{aa}

\usepackage{graphicx}
\usepackage{txfonts}

\usepackage[dvipsnames]{xcolor}
\usepackage{subcaption}
\usepackage{stfloats}
\usepackage{placeins}

\usepackage[breaklinks=true]{hyperref} 
\hypersetup{
	colorlinks = true, 
	urlcolor = blue,  
	linkcolor = blue, 
	citecolor = blue 
}
\usepackage{mathastext}
\newcommand{\orcid}[1]{\href{https://orcid.org/#1}{\includegraphics[scale=0.3]{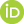}\,}}

\newcommand{\oiiau}{[\ion{O}{ii}]$\lambda7320,7330$} 
\newcommand{\hbl}{H$\beta$$\lambda4861$} 
\newcommand{\oiiibl}{[\ion{O}{iii}]$\lambda4959$} 
\newcommand{\oiiirl}{[\ion{O}{iii}]$\lambda5007$} 
\newcommand{\oiiit}{[\ion{O}{iii}]4959,5007}

\newcommand{\oit}{[\ion{O}{i}]$\lambda6300,6363$}
 
\newcommand{\niibl}{[\ion{N}{ii}]$\lambda6548$} 
\newcommand{\niirl}{[\ion{N}{ii}]$\lambda6584$} 
\newcommand{\niit}{[\ion{N}{ii}]$\lambda6548, 6584$} 
\newcommand{\niiau}{[\ion{N}{ii}]$\lambda5755$}
\newcommand{\nisky}{[\ion{N}{i}]$\lambda5755$}
\newcommand{\siibl}{[\ion{S}{ii}]$\lambda6716$} 
\newcommand{\siirl}{[\ion{S}{ii}]$\lambda6731$} 
\newcommand{\siit}{[\ion{S}{ii}]$\lambda6716,6731$} 
\newcommand{\siiiau}{[\ion{S}{iii}]$\lambda6312$} 
\newcommand{\siiil}{[\ion{S}{iii}]$\lambda9069$} 
\newcommand{\hb}{H$\beta$} 
\newcommand{\hei}{\ion{He}{i}} 
\newcommand{\heia}{\ion{He}{i}$\lambda5015$} 
\newcommand{\heib}{\ion{He}{i}$\lambda5875$} 
\newcommand{\heic}{\ion{He}{i}$\lambda6678$} 
\newcommand{\heid}{\ion{He}{i}$\lambda7065$} 
\newcommand{\ariii}{[\ion{Ar}{iii}]} 
\newcommand{\ariiib}{[\ion{Ar}{iii}]$\lambda7135$} 
\newcommand{\ariiir}{[\ion{Ar}{iii}]$\lambda7751$} 
\newcommand{\paninel}{Pa$9$$\lambda9229$} 
\newcommand{\patenl}{Pa$10$$\lambda9015$} 
\newcommand{\paelevenl}{Pa$11$$\lambda8862$} 
\newcommand{\patwelvel}{Pa$12$$\lambda8750$} 
\newcommand{\heii}{\ion{He}{ii}$\lambda4868$} 
\newcommand{\oiii}{[\ion{O}{iii}]} 
\newcommand{\oi}{[\ion{O}{i}]} 
\newcommand{\ha}{H$\alpha$} 
\newcommand{\nii}{[\ion{N}{ii}]} 
\newcommand{\sii}{[\ion{S}{ii}]} 
\newcommand{\siii}{[\ion{S}{iii}]} 
\newcommand{\hii}{\ion{H}{ii}}

\begin{document}

\title{M$^3$D: Mosaicking M33 with MUSE datacubes}

\subtitle{I. Unveiling the diversity of \hii\ regions in M33 with MUSE}

\author{A. Feltre
          \inst{\orcid{0000-0001-6865-2871}  1}\fnmsep\thanks{\email{anna.feltre@inaf.it}} 
          \and
          F. Belfiore\inst{\orcid{0000-0002-2545-5752} 1, 2}
          \and
          G. Cresci\inst{1}
          \and
          E. Corbelli\inst{1}
          \and
          N. Tomi\v{c}i\'{c} \inst{\orcid{0000-0002-8238-9210} 3}
          \and
          F. Mannucci\inst{1}
          \and 
          A. Marconi\inst{4,1}
          \and 
          E. Bertola\inst{1}
          \and 
          C. Bracci \inst{4,1}
          \and 
          E. Cataldi \inst{4,1}
          \and
          M. Ceci\inst{4,1}
          \and 
          M. Curti\inst{2}
          \and 
          Q. D'Amato\inst{1}
          \and
          M. Ginolfi\inst{4,1}
          \and 
          E. Koch \inst{5,6}
          \and 
          I. Lamperti\inst{3,1}
          \and 
          L. Magrini\inst{1}
          \and 
          C. Marconcini\inst{3,1}
          \and 
          A. Plat\inst{7}
          \and 
          M. Scialpi\inst{8,4,1}
          \and
          G. Tozzi\inst{9}
          \and
          L. Ulivi\inst{8,4,1}
          \and 
          G. Venturi\inst{10,1}
          \and 
          M.V. Zanchettin\inst{1}
          \and
          A. Chakraborty\inst{1}
          \and
          A. Amiri\inst{\orcid{0000-0002-8553-1964} 11}
           }
          
\institute{INAF - Osservatorio Astrofisico di Arcetri, Largo E. Fermi 5, I-50125, Florence, Italy\label{arcetri}
    \and 
    European Southern Observatory, Karl-Schwarzschild-Str. 2, 85748, Garching bei München, Germany \label{ESO}
     \and 
    Department of Physics, Faculty of Science, University of Zagreb, Bijeni\v{c}ka Cesta 32, 10000 Zagreb, Croatia
    \and 
      Dipartimento di Fisica e Astronomia, Università di Firenze, Via G. Sansone 1, I-50019, Sesto F.no (Firenze), Italy
    \and 
    Center for Astrophysics | Harvard \& Smithsonian, 60 Garden Street, 02138 Cambridge, MA, USA 
    \and
    National Radio Astronomy Observatory, 800 Bradbury SE, Suite 235, Albuquerque, NM 87106, USA
    \and
    Institute of Physics, GalSpec Laboratory, Ecole Polytechnique Federale de Lausanne, Observatoire de Sauverny, Chemin Pegasi 51, 1290 Versoix, Switzerland
    \and 
    University of Trento, Via Sommarive 14, Trento, I-38123, Italy
    \and
    Max-Planck-Institut f\"ur extraterrestrische Physik (MPE), Gie\ss enbachstraße 1, 85748 Garching, Germany
    \and 
    Scuola Normale Superiore, Piazza dei Cavalieri 7, I-56126 Pisa, Italy
    \and 
    Department of Physics, University of Arkansas, 226 Physics Building, 825 West Dickson Street, Fayetteville, AR 72701, USA
    }

   \date{}

\abstract{
We present new VLT/MUSE observations of a 3 $\times$ 8 arcmin$^2$ mosaic along the southern major axis of the Local Group galaxy M33. These data provide an unprecedented view of the galaxy's interstellar medium (ISM) and allow us to resolve ionised nebulae at a spatial scale of $\approx$5 pc. We identified and catalogued 131 \hii\ regions, down to \ha\ luminosities of $\approx 5\times$10$^{35}$ erg s$^{-1}$, one order of magnitude fainter than previous surveys on nearby galaxies beyond the Local Group, and we compared these regions with the spatial distribution of ionising stars and embedded star clusters. For each region, we extracted the corresponding integrated optical spectra and measured the intensity of key optical emission lines (\hbl, \oiiit, \niit, \ha, \siit, \siiil), other weaker optical lines when detectable, and Paschen lines to characterise the physical properties of the ioinised gas, such as density, dust attenuation, and metallicity. Our spatially resolved line ratio and flux maps reveal a remarkable diversity in ionisation properties, from dust-obscured regions hosting young stellar objects to highly ionised bubbles exhibiting high \oiii/\hb\ ratios. Our data reveal a diversity of ionisation fronts, ranging from well-defined to partial to absent. The radial profiles we obtained indicate the presence of both optically thin (density-bounded) \hii\ regions permitting the escape of ionising photons and fully ionised, optically thick (ionisation-bounded) \hii\ regions. The richness of this MUSE mosaic offers an unprecedented view of the ionised ISM at $\approx$5 pc resolution, providing direct insight into how stellar feedback shapes its environment. 
}

\titlerunning{Nebulae and ionising sources in M33}
\authorrunning{A. Feltre}
\maketitle
%
\section{Introduction}

The emission from ionised nebulae within galaxies encodes crucial information on the physical conditions of the gas and the nature of the ionising sources. 
As an example, nebular spectra from \hii\ regions carry information about the underlying ionising sources, such as massive stars and young stellar clusters. These stellar sources influence the local environment of \hii\ regions through radiation, stellar winds, and supernova explosions, collectively referred to as radiative (radiation) and mechanical (stellar winds and supernovae) feedback \citep{Pellegrini2011, Lopez2014, McLeod2019, McLeod2021, Kruijssen2019,  Olivier2021, Barnes2020, Barnes2021, Chevance2022}. By interpreting key optical emission-line ratios (e.g. \oiii/\hb, \nii/\ha, \sii/\ha), one can infer the physical and chemical conditions within \hii\ regions across a wide range of environments \citep[e.g.][and references therein]{Maiolino2019,Kewley2019}. Moreover, \hii\ regions dominate line emission in integrated galaxy spectra, providing a crucial observational link between the physical properties of individual massive stars (and young stellar clusters) and the global spectroscopic features of the nebular regions themselves \citep[e.g.][]{Kewley2002, Dopita2006, Gutkin2016}.

Traditionally, photoionisation models for \hii\ regions have often adopted simplified geometries (spherical or plane-parallel) and homogeneous density distributions. 
However, high-spatial resolution observations using integral field spectrographs (IFS), such as the Multi Unit Spectroscopic Explorer \citep[MUSE;][]{Bacon2010} at the Very Large Telescope (VLT) and the optical imaging Fourier transform spectrometer (IFTS) SITELLE \citep{Drissen2019} at the Canada France Hawaii Telescope (CFHT) have revealed a far larger degree of structural complexity. This includes \hii\ regions with diverse morphologies, internal gradients in ionisation and physical conditions, and porous boundaries that allow ionising radiation to leak into the surrounding ISM \citep[e.g.][]{Pellegrini2012, DellaBruna2021, DellaBruna2022a, Micheva2022, Mayya2023, Jin2023}. In parallel, theoretical models have evolved to incorporate these new observational constraints. Recent theoretical efforts have begun to explore more complex scenarios, incorporating inhomogeneous density distributions, radiative feedback, and evolving stellar populations within dynamical ISM environments (e.g. \citealt{Pellegrini2020, Jin2022, Jin2022b}).
Yet, bridging the gap between idealised models and the observed complexity requires large samples of spatially resolved nebulae, observed at sufficiently high spatial resolution to probe small scales (< 10 pc) and with a high signal-to-noise ratio (S/N).

In recent years, large surveys of nearby galaxies have contributed to advancing our view of star-forming regions and stellar feedback at small scales. Notable efforts include IFS observations, such as the MUSE Atlas of Discs (MAD) survey \citep{Erroz-Ferrer2017, denBrok2020} probing scales of $~$100 pc across a sample of 45 disc galaxies at an average distance of 20 Mpc, and the PHANGS-MUSE survey \citep{Emsellem2022}, which targeted 19 galaxies with MUSE at a distance of $\sim$14 Mpc, reaching scales of $\sim$40 pc. The SIGNALS survey \citep{Rousseau-Nepton2019} observed \hii\ regions across the disc of $\sim$40 galaxies, including M33 \citep{Rhea2020,Rhea2021}, with SITELLE on CFHT. Individual systems have been studied at even higher physical resolutions ($\sim$10 pc), approaching the scales needed to resolve the internal structure of \hii\ regions \citep[e.g. ][]{Tomicic2017, Tomicic2019, DellaBruna2020, DellaBruna2021,DellaBruna2022a, DellaBruna2022b, McLeod2020, Micheva2022, Congiu2025}. However, achieving meaningful advancements requires assembling a sufficiently large sample of \hii\ regions from observations able to resolve even smaller scales. Such resolutions can only be achieved by new-generation instruments, such as the Local Volume Mapper, part of SDSS V \citep{Drory2024, Kreckel2024}, for the Milky Way and the Magellanic Clouds. 

The Local Group galaxy M33 \citep[D$\sim$840 kpc, scale $= 4.0$ pc/\arcsec][]{Freedman1991, Gieren2013} is the only face-on actively star-forming spiral galaxy where we can map the nebular emission from different sources at the exquisite resolution of $\approx$5 pc with MUSE across a variety of environments, extending from the centre to the lower-metallicity galaxy outskirts. In contrast, the more evolved Sb galaxy M31 has a lower surface density of star formation than M33, which is a pure blue disc with no recent mergers. In addition, M31 has a high inclination angle ($\sim70^{\circ}$), which can hinder the interpretation of spatially resolved structures. These considerations make M33 a more favourable laboratory for such studies. 
In this paper, we analyse VLT/MUSE observations of a 3 $\times$ 8 arcmin$^2$ region of the southern major axis of M33 to trace ionised gas emission on scales of $\sim$ 5~pc. This data was exploited for the first time in \cite{Bracci2025}, who built a machine-learning powered classification model aimed at separating different ionised nebulae within the MUSE footprint.

The unique availability of multi-band datasets for M33 enables one to relate the energy sources of stellar feedback acting on parsec scales to the properties of the ionised gas in \hii\ regions, as well as the surrounding less-dense and warmer-diffuse ionised gas (DIG; T$\sim10^{4}$~K). The DIG can contribute up to 30–60\% of the total H$\alpha$ emission in local galaxies \citep[e.g.][]{Hoopes2000, Madsen2006, Haffner2009}, making it a crucial component to study in the context of feedback and ionising photon escape. Specifically, several complementary datasets are available for M33, including the PHATTER (UV to near-IR) Hubble Space Telescop (HST) imaging program \citep{Williams2021} and information on the cold ISM phases provided by \ion{H}{i} and CO maps \citep{Corbelli2014, Druard2014, Tabatabaei2014,  koch2018, koch2025, Muraoka2023,  Sarbadhicary2023}. In addition, JWST-MIRI data provide a direct, high-resolution view of the dusty ISM and embedded massive young stellar objects \citep[YSOs;][]{Peltonen2024}. These datasets are further complemented by catalogues of key stellar populations tracing different evolutionary stages, including young stellar cluster candidates \citep[YSCCs;][]{Corbelli2017}, star clusters from the PHATTER survey \citep{Johnson2022}, AGB stars \citep{McQuinn2007} and Wolf-Rayet (WR) stars \citep{Neugent2011}.

In this work, we present the MUSE mosaic observations of M33 (Sect. \ref{sec:data}), focusing on the identification and characterisation of \hii\ regions across the field. In Sect. \ref{sec:int}, we describe our method for identifying nebular regions and isolating \hii\ regions from supernova remnants (SNRs) and planetary nebulae (PNe). We also detail the 1D  spectral extraction and the integrated properties of these regions. 
Section \ref{sec:spatial} is dedicated to the spatial analysis of the nebular emission, while Sect. \ref{sec:radial_tot} is centred on the radial profiles of different emission line ratios. Finally, in Sect. \ref{sec:discussion}, we discuss our findings for specific peculiar \hii\ regions and compare the variety of nebular properties showcased by our data with theoretical predictions, before concluding in Sect. \ref{sec:conclusions}. 
\section{MUSE data of M33}\label{sec:data}

\subsection{Observations}

\begin{figure}[!b]
\includegraphics[width=0.5\textwidth, trim=10 0 170 30, clip]{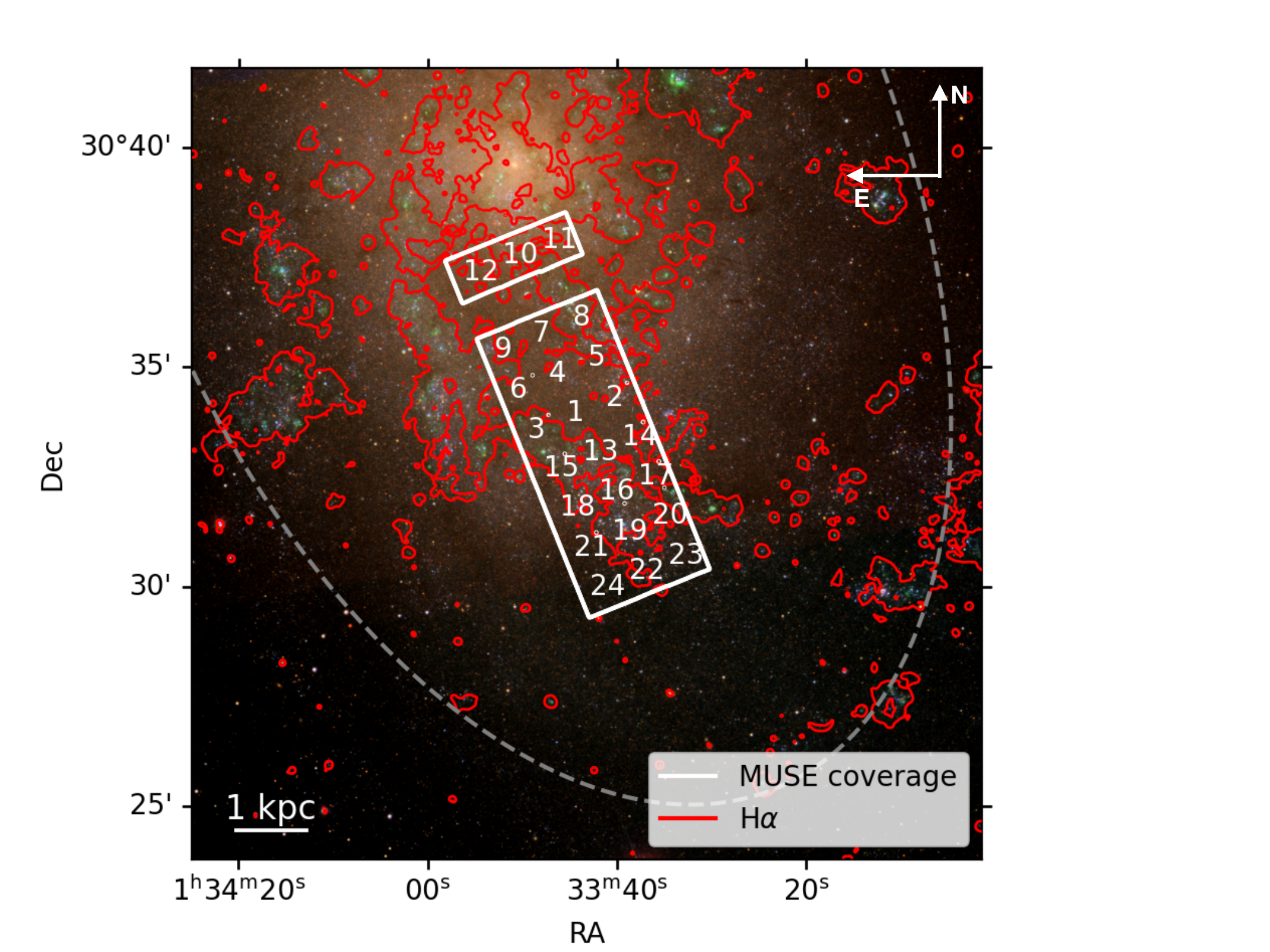}

\caption{Image of M33 using $g-r-i$ SDSS bands and showing the area of the galaxy targeted by MUSE (white contours). The positions of the centres of the individual MUSE pointings (1-24) are also shown. The red-coloured contours correspond to areas of high H$\alpha$ emission from the ground-based image of \cite{Massey2006}.}
    \label{fig:footprint}
\end{figure}

Optical integral field spectroscopy was acquired with the MUSE instrument \citep{Bacon2010} on the VLT UT4 as part of the program ID 109.22XS.001 (PI: G. Cresci). MUSE was used in its seeing-limited, wide-field mode with a nominal wavelength range of 4750-9350~$\AA$. In this configuration, a MUSE pointing covers a field of view of 1 $\times$ 1 arcmin$^2$, with a pixel scale of 0.2 arcsec and a spectral resolution of R $\sim$ 3000. In our observations, we adopted 24 individual pointings to cover a mostly contiguous region of 3 $\times$ 8 arcmin$^2$, along the southern major axis of M33 (see Fig. \ref{fig:footprint}), with an overlap of 2 arcsec between contiguous pointings.

We observed three object pointings per observing block (OB), each making up small strips, oriented 22$^\circ$ east of north. Each pointing in the strip was observed with two consecutive exposures of 340 seconds, with the second exposure observed after a 90$^\circ$ rotation of the instrument. Since all our object pointings are entirely within M33, to obtain accurate sky subtraction, after each object OB we observed a separate sky OB for 120 seconds, targeting an emission-free area sufficiently far from the galaxy.

Observations were taken between September 5 and October 4, 2022. Each OB, corresponding to a 3 arcmin strip, was observed in different weather conditions (see Table \ref{table:obs}). All observations were obtained with seeing better than 1\arcsec\ (mostly 0.6\arcsec--0.8\arcsec), except for the OB of pointings 7-9 (Table \ref{table:obs}) which was observed under significantly worse seeing (1.5\arcsec). An observing log is provided in Appendix \ref{app:AppA}. The gap in the northern part of the MUSE footprint (between pointings 10-12 and 7-9) was caused by technical issues during observations.

\subsection{Data reduction}

The MUSE data reduction was performed with the {\tt pymusepipe} software package\footnote{\url{https://pypi.org/project/pymusepipe/}}, a customised Python wrapper to the MUSE data reduction pipeline \citep{Weilbacher2020} developed by the PHANGS team \citep{Emsellem2022} and optimised for processing large mosaics of extended objects. For each OB, {\tt pymusepipe} implements the standard data reduction steps by running the MUSE recipes provided by {\tt esorex}, including bias subtraction, flat fielding, wavelength calibration, line spread function computation, illumination correction, reduction of the standard stars, and extraction of the sky spectrum from the dedicated sky exposures. In addition, {\tt pymusepipe} automatically matches object exposures with associated calibrations, preferring calibration files which are closest in time. 

For each individual exposure, a fully reduced preliminary datacube is generated and collapsed to generate an $r$-band image. The individual $r$-band images of the 48 MUSE exposures (two exposures per each of the 24 pointings) are then astrometrised with respect to a background mosaic of the same field as observed by the Sloan Digital Sky Survey (SDSS) \footnote{\url{https://www.sdss.org/}} in the same band (Fig \ref{fig:footprint}). This alignment step is performed using {\tt spacepylot}\footnote{\url{https://github.com/ejwatkins-astro/spacepylot}}, a routine optimised for finding shifts between astronomical images containing extended emission and based on the optical flow algorithm. Specifically, {\tt spacepylot} automatically derives both translational and rotational offsets between the individual exposures and the SDSS mosaic and is natively integrated in the {\tt pymusepipe} workflow. In a few cases the offsets produced by {\tt spacepylot} were not optimal and they were further tweaked manually using a graphical interface provided by {\tt pymusepipe}. We find that the relative offsets between individual exposures of the same field are generally comparable with the MUSE pixel scale (0.2$''$), but absolute offsets with respect to SDSS were significant before the astrometric correction (median of 1.5$''$ and up to 2.5$''$). 

Following \cite{Emsellem2022}, we also use the SDSS mosaic to fix the absolute flux calibration of the MUSE data. In particular, after fixing the absolute astrometry of each exposure we calculate a linear regression between the SDSS and MUSE reconstructed $r$-band flux in bins of $15\times15$ spaxels ($3 \times 3$ arcsec$^2$). The slope of this relation is interpreted as a multiplicative correction to the absolute flux calibration of the MUSE data, and the intercept as a residual background term due to imperfect sky subtraction. Finally, the astrometric offsets in RA and DEC and the multiplicative flux calibration term are recorded in a table used by the MUSE reduction pipeline. Any applied rotation or background term are also recorded and applied as part of the next step of the {\tt pymusepipe} processing. 
The final product of the data reduction is a fully reduced mosaic containing all the MUSE exposures.

\subsection{Spectral fitting and emission lines mapping}
\label{sec:linemaps}

We derived emission-line maps from the final datacubes using the data analysis pipeline\footnote{\url{https://gitlab.com/francbelf/ifu-pipeline/}} developed for the PHANGS project. For this work we performed two subsequent spectral fitting steps. The first was performed on a Voronoi binned \citep{Cappellari2003} version of the data with a target signal-to-noise ratio of 35 in the continuum region around 5400~\AA\ free of emission lines. This procedure led to convex bins with a median circularised radius of $\sim$ 0.8$''$. The stellar continuum is then fitted with a set of eMILES simple stellar population models \citep{Vazdekis2016}, while masking the emission lines. The fit is performed with the penalised pixel fitting (pPXF) python routine described in \cite{Cappellari2004} and \cite{Cappellari2017}.  

In the second fit iteration we fit the spectra at the level of single spaxels and simultaneously fit the stellar continuum and the emission lines via a new call to pPXF. The stellar kinematic parameters (velocity and velocity dispersion) obtained in the previous fit iteration are then fixed for this second iteration, which is optimised for the extraction of emission-line parameters. We fit the major emission lines in the MUSE wavelength range, including \hbl, \oiiit, \niit, \ha, \siit\ and \siiil, with Gaussian functions and tie the kinematic parameters in three groups: Balmer lines, low-ionisation and high-ionisation lines. Lines whose relative flux ratio is fixed by atomic physics (e.g. \oiiirl/\oiiibl $\approx 2.98$ and \niirl/\niibl $\approx 2.96$) are given additional constraints on their amplitudes. These choices largely mirror those made in previous work and are extensively motivated therein \citep{Belfiore2019, Emsellem2022}. The pipeline produces output files containing stellar kinematics, fluxes and kinematic parameters of the fitted emission lines in each spaxel. All line fluxes are corrected for Milky Way foreground extinction \citep[details in Sect. 2.2.3 of][]{Belfiore2019}.

\begin{figure*}[p]
\centering
\includegraphics[width=0.95\textwidth]{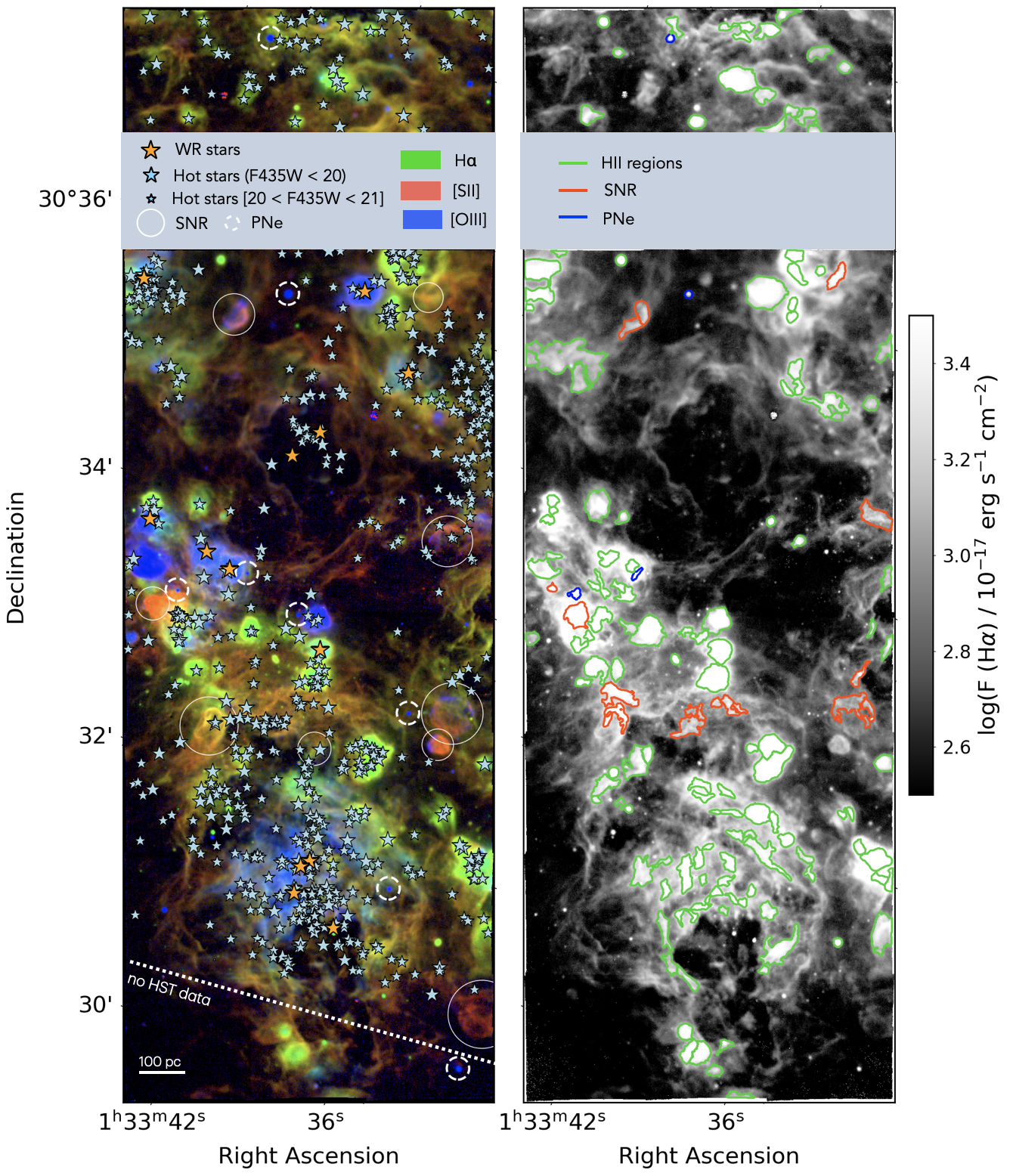}
\caption{Emission maps of M33 MUSE data. Left: three-colour image from MUSE mapping the ionised gas emission: \oiiirl\ in blue, \ha\ in green and \siit\ in red. Light-blue stars are the 552 stars selected from the PHATTER survey \citep{Williams2021} as described in Appendix \ref{sec:phatter}. Smaller and larger star symbols denote two ranges of F475W magnitude, between 20 and 21 and below 20, respectively. Orange stars are WR stars from \cite{Neugent2011,Neugent2019} (Appendix \ref{sec:wr}). White solid and dashed circles denote SNRs from \cite{Lee2014} and PNe from \cite{Ciardullo2004}, respectively (Appendix \ref{sec:snr} and \ref{sec:pne}). Right: \ha\ map from MUSE with the final configuration of the nebular masks, colour-coded on the basis of their classification as labelled in the legend.}
    \label{fig:M33_muse}
\end{figure*}

A three-colour image of \ha\ (green), \oiiirl\ (blue), and \siit\ (red) line fluxes from the MUSE data is shown in Fig. \ref{fig:M33_muse} (left panel). 
Broadly, our data covers two spiral arms, crossing the image from the top-left to bottom-right, and an extended patch of the inter-arm region. Details on the ancillary data used in this work to trace the different ionising sources are provided in Appendices \ref{app:AppB}.

\clearpage
\section{Integrated properties of \hii regions} \label{sec:int}

\subsection{Detection and segmentation of ionised nebulae}
\label{sec:astrodendro}

The two spiral arms in Fig. \ref{fig:M33_muse} show extensive line emission, including classical \hii\ regions powered by massive stars, PNe (white dashed circles), and SNRs (white circles). 
We identified the \hii\ regions using a two-step procedure. First, we employed the Python package \texttt{astrodendro}\footnote{\url{https://dendrograms.readthedocs.io}} \citep{Robitaille2019}, which organises the input data into a hierarchical tree structure consisting of leaves and branches, to distinguish areas of nebular emission from the background. We provided the combined \ha$+$\oiii\ emission map as input to \texttt{astrodendro}, instead of the \ha\ map only, as we found that this choice maximises the identification of ionised nebulae while minimising the inclusion of \ha-emitting filamentary structures and DIG regions. We tuned the parameters of \texttt{astrodendro} to reduce spurious detections and set \texttt{min\_value}=25$\sigma$, where $\sigma$ is the median error of the \ha$+$\oiii\ map and corresponds to a flux of 5.1$ \times 10^{-19}$ erg s$^{-1}$ cm$^{-2}$. We adopted the same value for the minimum difference between two separate structures (\texttt{min\_delta}), and required a minimum number (\texttt{min\_npix}) of 250 pixels per leaf. With a spatial resolution of MUSE having a full width at half maximum of $\approx$2.5 pixels, this corresponds to $\approx$40 resolution elements per leaf.

In a second step, we applied a watershed segmentation algorithm \citep{Vincent1991} to the regions defined by \texttt{astrodendro} to separate individual emission regions. This algorithm treats the intensity map as a topographic surface and separates blended regions by assigning pixels to local maxima based on intensity gradients. In particular, we applied the \texttt{deblend\_sources} function from the \texttt{photutils} package. A low contrast threshold of 0.001 was used to identify faint overlapping sources, and a minimum pixel requirement of 150 was set to avoid segmenting noise or spurious detections. The total number of nebular emission regions identified with our two-step procedure is 153 (Fig. \ref{fig:M33_muse}, right panel).

We extracted the integrated spectra of these 153 nebular regions by summing the flux within the masked areas defined as described above. 
Following the same procedure described in Sect. \ref{sec:linemaps}, we fitted several emission lines, including strong optical lines (\hbl, \oiiit, \niit, \ha, \siit\ and \siiil), \hei\ lines (\heia, \heib, \heic, \heid), auroral lines (\niiau\ and \oiiau), \ariii\ lines (\ariiib\ and \ariiir) and lines of the Hydrogen Paschen series (\patwelvel, \paelevenl, \patenl, \paninel).
We do not have access to accurate measurements of \nisky, \oit\ and \siiiau\ because of strong contamination of skyline residuals, and therefore we did not include these lines in the fitting. 

We have released the line measurements for all the ionised nebulae classified as \hii\ regions (see Sect. \ref{sec:class_hii}), along with other properties computed on their integrated spectra (see Sect. \ref{sec:hii_int}). Details on the content of this catalogue are provided in Appendix \ref{app:AppD} and the catalogue is available at the CDS. We do not include PNe and SNRs, as the statistics are too limited compared with literature catalogues, and our nebular masks were defined to minimise their contribution.

\subsection{Classification of emission line nebulae}
\label{sec:class_hii}

We used literature catalogues (Appendix \ref{app:AppB}) and measured line ratios to classify \hii\ regions and distinguish them from PNe and SNRs.
Four regions (blue contours in Fig. \ref{fig:M33_muse}, right panel) are classified as PNe by \citet[][see Appendix \ref{sec:pne}]{Ciardullo2004}. In addition, thirteen regions (red contours) either fully or partially overlapping with the SNR from the catalogue of \cite{Lee2014}. The line ratios measured on the MUSE spectra of these regions confirm the same classification. 

We also looked for additional PNe or SNR sources in our data, inspecting those nebulae whose \oiii/\hb\ (high in PNe) and \sii/\ha\ (high in SNR) ratios are above the 95 percentile of the line-ratio distributions of the entire sample of nebular regions.
Among the dendrograms with the highest \oiii/\hb\ ratio not included in previous PNe catalogues, three host WR stars classified by \cite{Neugent2011}. The other three regions exhibit extended \oiii\ confirming their \hii\ region nature, but suggestive of a hard ionising field or a nebular geometry consistent with a density-bounded region, allowing the escape of ionising photons (see Sects. \ref{sec:neb_geo} and \ref{sec:radial_tot}).
Among the dendrograms with the highest \sii/\ha, five regions exhibit very elongated nebular geometries with irregular and jagged contours. We flagged these regions as potentially shocked regions, also shown with red contours in Fig. \ref{fig:M33_muse}. After removing these PN and SNR candidates, the final number of emission nebulae identified in Sect. \ref{sec:astrodendro} classified as \hii\ regions is 131.

The identified \hii\ regions exhibit a wide variety of nebular morphologies. While some conform to the canonical idealised definition of \hii\ region, with a single ionising star at the centre, others host multiple stars or even star clusters (as can be seen by comparing the ionising sources and the nebular masks in the left and right panels of Fig. \ref{fig:M33_muse}, respectively). 
In the following, we analyse all the 131 nebular regions, irrespectively from the number of ionising sources hosted within them.

\begin{figure}
\centering
\includegraphics[width=0.4\textwidth]{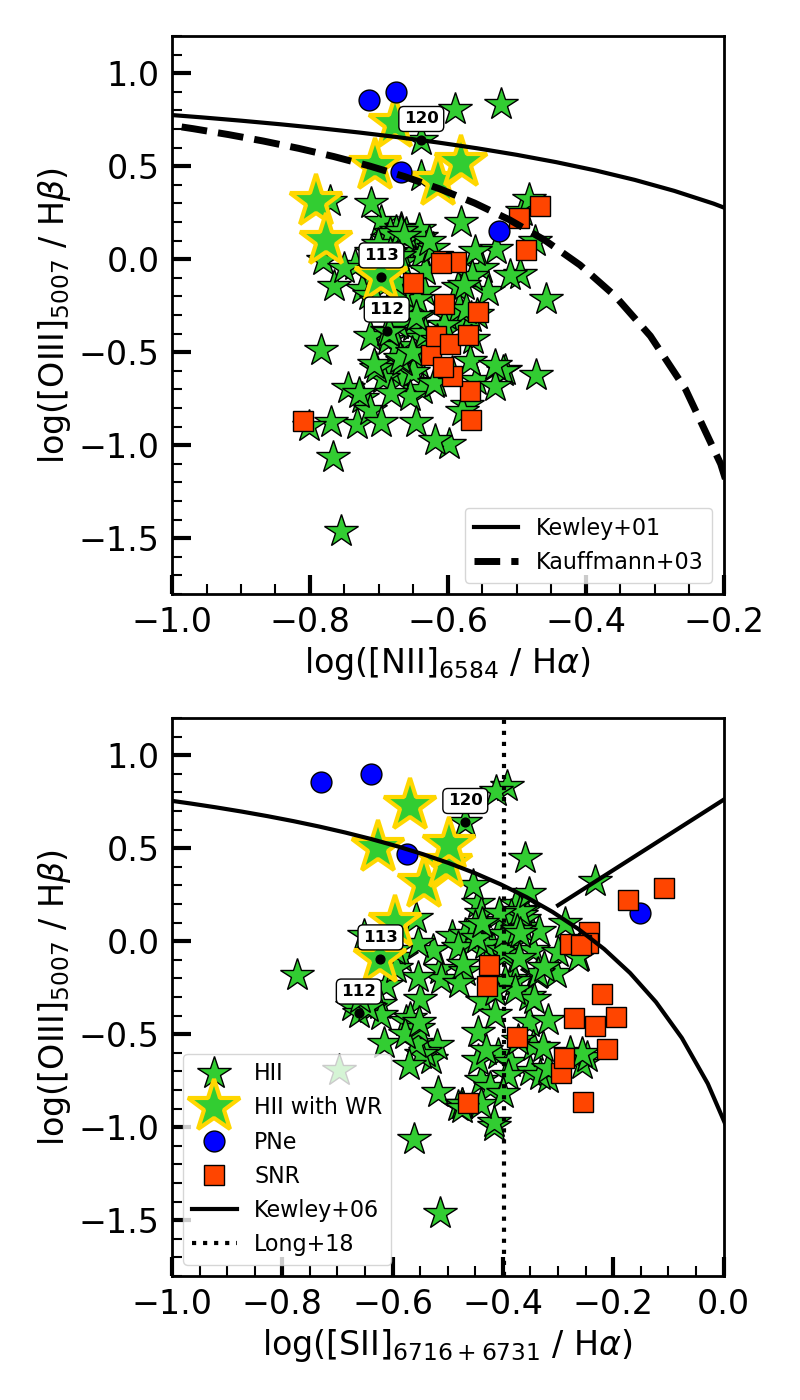}
  \caption{Optical  \oiii/\hb\ versus \nii/\ha\ (top) and \sii/\ha\ (bottom) diagnostic diagrams from \cite{Baldwin1981} and \cite{Veilleux1987}, respectively. Continuous and dashed curves show the demarcation criteria by \cite{Kewley2001, Kewley2006} and \cite{Kauffmann2003}, respectively. The dotted line indicates the traditional \sii/\ha\ cutoff of 0.4 for SNR \citep[e.g.][]{Long2018}. Green stars are the integrated measurements of \hii\ regions, blue circles of PNe and red squares of SNRs as labelled in the legend. The \hii\ regions with yellow edges host WR stars. Black dots indicate the position in the diagram of three \hii\ regions (IDs\, 112, 113 and 120) which are discussed in more details in Sect. \ref{sec:neb_geo}.}
        \label{fig:diagnostics}
  \end{figure}

Figure \ref{fig:diagnostics} illustrates how the different classes of ionised nebulae populate classical ionisation diagnostic diagrams. Specifically, the two panels show the \oiiirl/\hb\ versus \niirl/\ha\ and (\siibl+\siirl)/\ha\ diagrams \citep{Baldwin1981, Veilleux1987}, commonly used to disinguish star-forming regions from other ionising sources (e.g. AGN, PNe). In these diagrams, most of the \hii\ regions lie below the demarcation lines of \cite{Kewley2001,Kewley2006} and \cite{Kauffmann2003} (continuous and dashed curves, respectively). The regions classified as SNR populate the right locus in the bottom diagram, which is in good agreement with the commonly used cutoff value of \sii/\ha $= 0.4$ for SNR \citep[e.g. ][]{Long2018}. As explained in Sect. \ref{sec:astrodendro}, the nebular masks were identified from peaks in the combined \ha$+$\oiii\ emission maps, effectively reducing contamination from DIG, which typically exhibits \sii/\ha\ ratios larger than 0.3.

\cite{Bracci2025} presented a neural network trained on simulated spectra to classify nebular regions, including the integrated spectra of M33 nebulae from the MUSE data presented here. Our classification, based on more standard diagnostics, broadly agrees with their results. 
In particular, there is an excellent agreement between the neural network classification and the standard diagnostics for PNe and \hii\ regions. \cite{Bracci2025} also find that the secure identification of SNRs is more challenging, including when using neural network classification. This is particularly  evident in the low surface brightness regime, as in this MUSE M33 data ($\approx 2 \times 10^{39}$ erg s$^{-1}$ kpc$^{-2}$). This is because SNRs could be confused with DIG or faint \hii\ regions. For example, regarding the five SNRs candidates mentioned above, \citet{Bracci2025} classify four of them (IDs\,105, 109, 110 and 111) as SNRs, while one (ID\,105) is identified as DIG emission. Furthermore, the regions exhibiting extended, extreme \oiii\ emission are consistently classified as PNe by the neural network, due to their high-ionisation spectral features resembling those of PNe.

\subsection{Emission properties of \hii\ regions}
\label{sec:obs_prop}

Our \hii\ regions have observed \ha\ luminosities ranging between $4.6 \times 10^{35}$ and $3.7 \times 10^{37}$ erg s$^{-1}$, range similar to that probed by \cite{Rogers2022} in their M33 data, reaching values approximately one order of magnitude fainter than the completeness limit of the \hii\ region samples identified in local galaxies observed in the PHANGS-MUSE survey \citep{Santoro2022}. The \ha\ surface brightness comprises values between 1.5 $\times \, 10^{39}$ and 2.2 $\times \, 10^{40}$ erg s$^{-1}$ kpc$^{-2}$, straddling the traditional dividing line between \hii\ regions and DIG \citep{Belfiore2022}.
The high spatial resolution of our MUSE data allows us to split large \hii\ regions into smaller components, whereas lower-resolution studies typically measure integrated emission over larger, blended areas. However, when we account for the contribution from multiple adjacent nebular masks, we find good agreement with the \ha\ luminosities reported by \cite{Corbelli2009}, \cite{Grossi2010}, \cite{Sharma2011}, and \cite{Relano2013}, at least for the limited number of \hii\ regions (fewer than 20) from these works that fall in the MUSE footprint. Remaining discrepancy with the MUSE measurements are likely due to the use of nebular masks instead of circular apertures, which may exclude low-surface-brightness emission, possibly associated with DIG, at the edges of the masks.
 
Our \hii\ regions span approximately two orders of magnitude in the \oiii/\hb\ ratio (Fig.~\ref{fig:diagnostics}), reflecting a wide range of ionisation parameters \citep{Gutkin2016}. The scatter in \nii/\ha\ is smaller due to its higher sensitivity to metallicity, which exhibits limited variations within our sample (see Sect. \ref{sec:hii_int}). As illustrative examples, we selected three \hii\ regions (IDs\,112, 113 and 120) that exhibit distinct line ratios in the BPT-diagrams \citep{Baldwin1981, Veilleux1987} of Fig.~\ref{fig:diagnostics}, likely due to different properties of the ionising sources (Sect. \ref{sec:class_hii}). 

The difference in nebular line ratios of ID\,120 with respect to ID\,112 and 113 is evident. The latter two exhibit line ratios in the areas occupied by classical \hii\ regions in the BPT diagrams, while ID\,120 occupies the region above the classical demarcation lines lines from \cite{Kewley2001}, typically associated to PNe and AGN.

Because of the hard ionisation field produced by WR stars \citep{Shirazi2012}, the seven \hii\ regions hosting a WR star have higher \oiii/\hb\ than the average of \hii\ regions, with integrated line ratios close to those of PNe (Fig. \ref{fig:diagnostics}).  Four of these WR-driven nebular regions, namely  ID\,40, ID\,50, ID\,113 and ID\,120, are part of the sample of \oiii\ bubbles surrounding WR stars identified by \cite{Tuquet2024} using the observations of M33 from the imaging Fourier transform spectrometer SITELLE \citep{Drissen2019}. 
However, our data shows that the strong \oiii\ emission is not exclusive to \hii\ regions hosting WR stars. In fact, other \hii\ regions, such as ID\,120, exhibit similar integrated line properties without hosting a WR star. 
To confirm this conclusion, we extracted the spectra of stars (HST passband F435W $<$ 21 mag) located within the three regions without catalogued WR stars that have the highest \oiii/\hb\ ratio (IDs\,120, 127, 128). In these spectra, we did not find signatures typical of WR stars, such as the 5800 $\AA\ $bump. These features are instead clearly visible in the MUSE spectra of WR stars selected from the \cite{Neugent2011} catalogue (Appendix \ref{sec:wr}), suggesting a different origin for the extreme line ratio observed in the regions mentioned above. We also do not detect clear spectroscopic signatures of other feedback-driving sources, such as O-type stars \citep{McLeod2020}, because of the intense nebular emission in which they are embedded, which hampers the extraction of clean stellar spectra. Examples of regions with intense and extended \oiii\ emission are presented in Fig. \ref{fig:oiii_bubbles} and further discussed in Sect. \ref{sec:oiii_bubbles}.

\subsection{Physical properties of \hii\ regions}

\label{sec:hii_int}

We computed the electron density from the \siibl\ to \siirl\ ratio via the \texttt{pyNeb} python package \citep{Luridiana2015} assuming an electron temperature T$_{\rm e}$ of 10$^4$ K. We observe electron densities below the low-density limit of $n_{\rm e} \sim 40$ cm$^{-3}$ (below which \sii6716/6730 ratio saturates) for nearly all of our \hii\ regions (97\% of the sample).

We computed dust reddening, E(B-V), from the Balmer decrement. We assume case B hydrogen recombination value for \ha/\hb\ \citep[i.e. 2.86 for an electron temperature of 10$^4$ K and an electron density of 10$^2$~cm$^{-3}$;][]{Hummer&storey1987} and the attenuation law of \cite{Odonnell1994}, appropriate for the optical emission from individual \hii\ regions in M33 and consistent with previous studies of its nebular dust properties. The value of the theoretical \ha/\hb\ ratio is only weakly dependent on the electron density within the low-density regime (< 10$^2$~cm$^{-3}$), and remains effectively unchanged for densities as low as those found in our sample. Most of our \hii\ regions have E(B-V) comprised between 0.05 and 0.60, and only two regions show higher attenuation (E(B-V) $>$ 0.3). The mean E(B-V) value is 0.15. 

We estimated the metallicity, expressed in terms of the oxygen abundance O/H, from the S-calibration presented in \citet[][see their Sect. 3.2.2]{Pilyugin2016} and from the N$_2$S$_2$H$\alpha$ calibrator presented in \cite{Brazzini2024}. In particular, the S-calibration from \cite{Pilyugin2016} uses the \oiiit/\hb, \niit/\hb\ and the \siit/\hb\ line ratios which are all in the MUSE spectral coverage. This empirical calibration was specifically derived for \hii\ regions and has the advantage of not being limited to a narrow range of metallicities. We use the S-calibration as the [\ion{O}{ii}]{3727,3729} line needed for the R-calibration, also presented in \cite{Pilyugin2016}, is not covered by the MUSE spectral range. The two calibrations, however, agree within $\sim$0.05 dex \citep{Pilyugin2016}.

\cite{Brazzini2024} derived updated calibrations using a sample of PHANGS-MUSE galaxies \citep{Groves2023} with auroral line detections available, complemented by additional auroral-line measurements from the literature. Among the indicators explored in \cite{Brazzini2024}, we employ here the N$_2$S$_2$H$\alpha$ diagnostic, as it exhibits the lower intrinsic scatter \citep[see table 5 of ][]{Brazzini2024} and avoids the double-branch behaviour characteristic of some other metallicity calibrators.

We measure a negative metallicity gradient, finding slopes of $-0.061$ and $-0.056$ dex kpc$^{-1}$ within the inner 2.5 kpc when using the S-calibration and the N$_2$S$_2$H$\alpha$ indicator, respectively. Our results confirm the negative gradients reported in previous studies \citep[e.g.][]{Rosolowsky2008, Magrini2010, Toribio2016, Rogers2022}.

The S-calibration yields systematically lower metallicity estimates compared to the N$_2$S$_2$H$\alpha$ diagnostic. This is consistent with the findings of \cite{Brazzini2024}, which show that the \citet{Pilyugin2016} calibration tends to produce lower metallicities than those obtained in their work, where metallicities were derived by minimising the $\chi^2$ defined simultaneously over all selected diagnostics (their equation 15).

\begin{figure}[!h]
\centering
\includegraphics[width=0.45\textwidth]{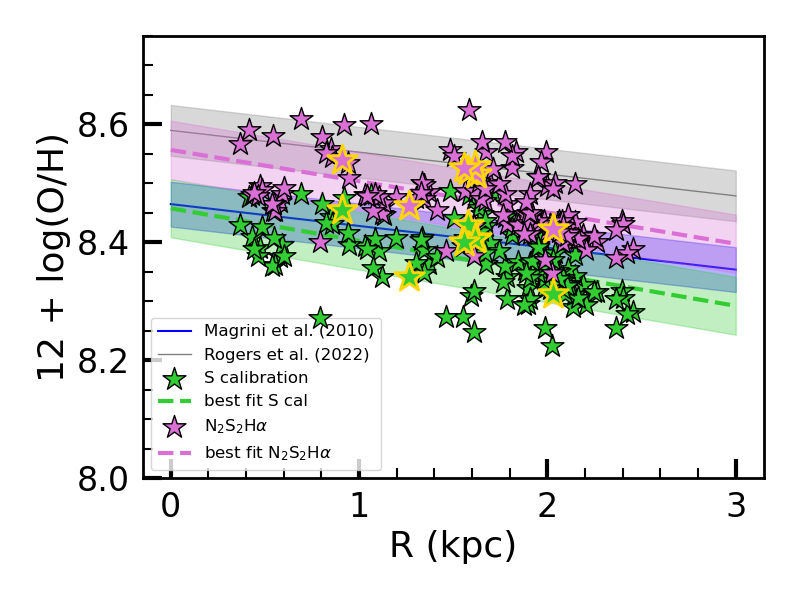}
  \caption{Metallicity gradients inferred from the \cite{Pilyugin2016} S-calibration (green) and the N$_2$S$_2$H$\alpha$ (violet) from \cite{Brazzini2024}. The dashed green and violet lines indicate the best-fits to our data, and the shaded area indicates the 1-$\sigma$ scatter. Blue and gray lines and shaded areas are the gradients obtained from direct oxygen abundance measurements by \cite{Magrini2010} and \cite{Rogers2022}, respectively. The stars yellow-edged stars are \hii\ regions hosting WR stars.}
        \label{fig:met_gradient}
  \end{figure}

 In Fig. \ref{fig:met_gradient}, we compare our results with the relation reported by \cite{Magrini2010} and \cite{Rogers2022}. 
Their work extends to galactocentric distances larger than 2.5 kpc (up to $\sim$ 8 kpc), exceeding the range covered by the MUSE mosaic, which may explain the observed flattening of the gradient in their study. 
The gradient obtained from the S-calibration, is consistent, and shows comparable scatter ($\approx 0.1$ dex), with that obtained by oxygen abundance determinations via the `direct method' \citep{Rosolowsky2008, Magrini2010, Toribio2016}, which constrains the electron temperature using the \oiii$ \lambda4363$ auroral line. The gradient obtained from N$_2$S$_2$H$\alpha$ is closer to that reported by \cite{Rogers2022} than the gradient derived from the S-calibration, which presented oxygen abundance determinations of $\sim$100 \hii\ regions in M33, observed with the Large Binocular Telescope (LBT) as part of the CHemical Abundances Of Spirals (CHAOS) project \citep{Berg2015}, using the weighted-average electron temperature prioritisation from \cite{Berg2020} and \cite{Rogers2021}. 
Their analysis attributes the offset and scatter among different literature results to the use of a single auroral line for electron temperature determination. They find that estimates obtained using multiple auroral lines reduce the inconsistencies among different measurements. In addition, aperture effects due to a slit width (1\arcsec) smaller than the typical size of the observed \hii\ regions can affect oxygen abundance determination and lead to discrepancies compared to other estimates \citep{Mannucci21}.

In Appendix \ref{app:AppC}, we report on weaker emission lines that are detected only in part of our dataset. These lines are also used to derive alternative diagnostics for electron density and dust attenuation, leading to consistent results to those presented in this section.

\section{Spatial structure of line emission}\label{sec:spatial}

\subsection{Line ratios in nebulae and diffuse regions}\label{sec:zoom-in}

\begin{figure*}
\centering
\includegraphics[width=0.99\textwidth, trim=20 30 20 10, clip ]{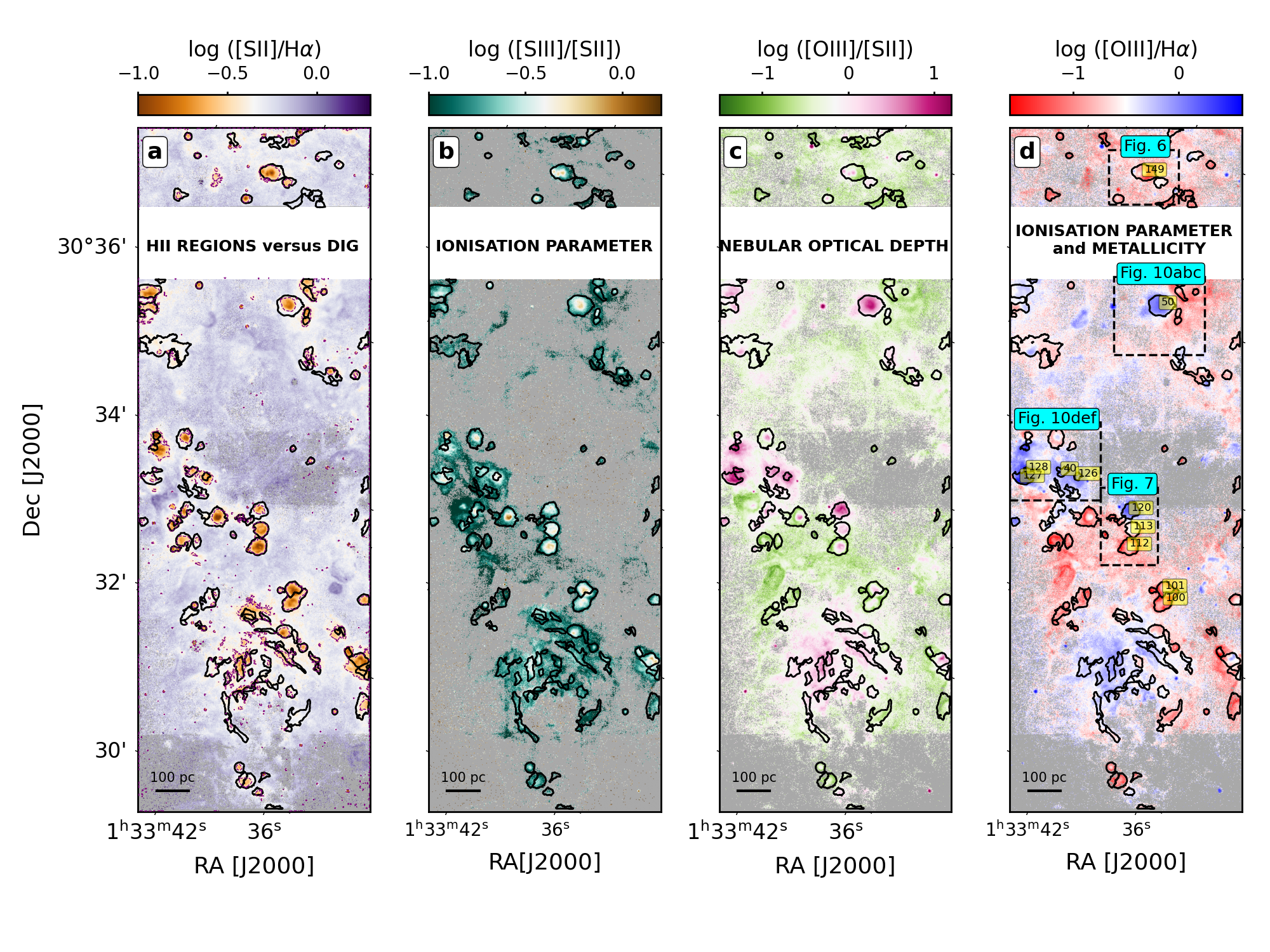}
  \caption{Line-ratio maps, corrected for dust attenuation, of our M33 data for a S/N of the lines of interest larger than 2.5. From left to right: \siit/\ha, \siiil/\siit, \oiiirl/\siit, and \oiiirl/\ha. Black contours indicate the dendrograms classified as \hii\ regions. In the \sii/\ha\ maps (panel a), the purple dashed contours show the level value of 0.35.}
        \label{fig:line_ratio_maps}
\end{figure*}

We examined several 2D line ratio maps, corrected for dust attenuation using the Balmer decrement and the \cite{Odonnell1994} attenuation curve, and identify nebular emission regions of particular interest. We applied a S/N cut of 2.5 to the lines involved in each ratio of interest.

In Fig. \ref{fig:line_ratio_maps}a, we show a map of \sii/\ha, which is commonly used for distinguishing the emission from the DIG and \hii\ regions due to its sensitivity to the ionisation state of the gas. The DIG around \hii\ regions is characterised by emission-line ratios that differ from classical \hii\ regions \citep{Haffner2009}, such as higher \sii/\ha\ and \nii/\ha\ \citep{Madsen2006, Zhang2017, Belfiore2022}.
An increase in \sii/\ha\ is observed in Fig. \ref{fig:line_ratio_maps}a moving from compact nebulae to more diffuse regions. For example, there is a good correspondence between the \hii\ regions defined via our nebular mask analysis and regions with \sii/\ha\ $<$ 0.35 (dashed violet contours in Fig. \ref{fig:line_ratio_maps}a). However, several nebulae defined from our dendrogram analysis show \sii/\ha\ up to $\approx0.5$ in at least part of the mask, pointing towards a non-negligible DIG contribution. 
Nevertheless, the demarcation between DIG and \hii\ regions varies depending on local gas physical conditions such as metallicity (still, the metallicity gradient is relatively flat in M33) and ionisation parameter. This challenges the use of a single diagnostic to separate DIG from \hii\ regions across different environments. The map in Fig. \ref{fig:line_ratio_maps}a demonstrates how spatially resolved data presented here can provide valuable insight in this context, allowing to study local variations in line ratios and ionisation conditions and helping to distinguish between diffuse and compact ionised gas structures.

Fig. \ref{fig:line_ratio_maps}b presents a map of the \siii/\sii\ ratio, which serves as a tracer of the ionisation parameter, and is largely independent of metallicity \citep[][]{Mathis1982, Mathis1985, Dopita1986, Kewley2002, Mingozzi2020}. This ratio and, analogously, the \oiii/\sii\ (Fig. \ref{fig:line_ratio_maps}c), are particularly effective at highlighting ionisation fronts. Most \hii\ regions exhibit high \siii/\sii\ and \oiii/\sii\ in their cores, with lower values (log(\siii/\sii) and log(\oiii/\sii) < -0.5) outlining ionisation fronts. 

Further insights into the nebular geometry and optical depth can be gained from the \oiii/\ha\ ratio (Fig. \ref{fig:line_ratio_maps}d) which, analogously to \oiii/\hb\ (not shown because of the lower S/N of \hb\ compared to \ha), provides information on the hardness of the ionising radiation and potential ionising photon leakage. While intense \oiii/\ha\ or, analogously, \oiii/\hb\ (log(\oiii/\ha) > 0.0 and  log(\oiii/\hb) > 0.6) are expected in WR-hosting regions (e.g. IDs\,40, 50, 113, 126, as labelled in the boxed areas of Fig. \ref{fig:line_ratio_maps}d), similarly high values are observed in \hii\ regions without WR stars (e.g. IDs\,120, 127,  labelled in Fig. \ref{fig:line_ratio_maps}d), suggesting alternative hard ionising sources. In some cases, the \oiii\ emission extends beyond the nebular mask, giving rise to larger structures that link multiple \hii\ regions. These structures are explored further in Sect. \ref{sec:oiii_bubbles}.

\begin{figure*}[!h]
\centering
\includegraphics[width=0.99\textwidth]{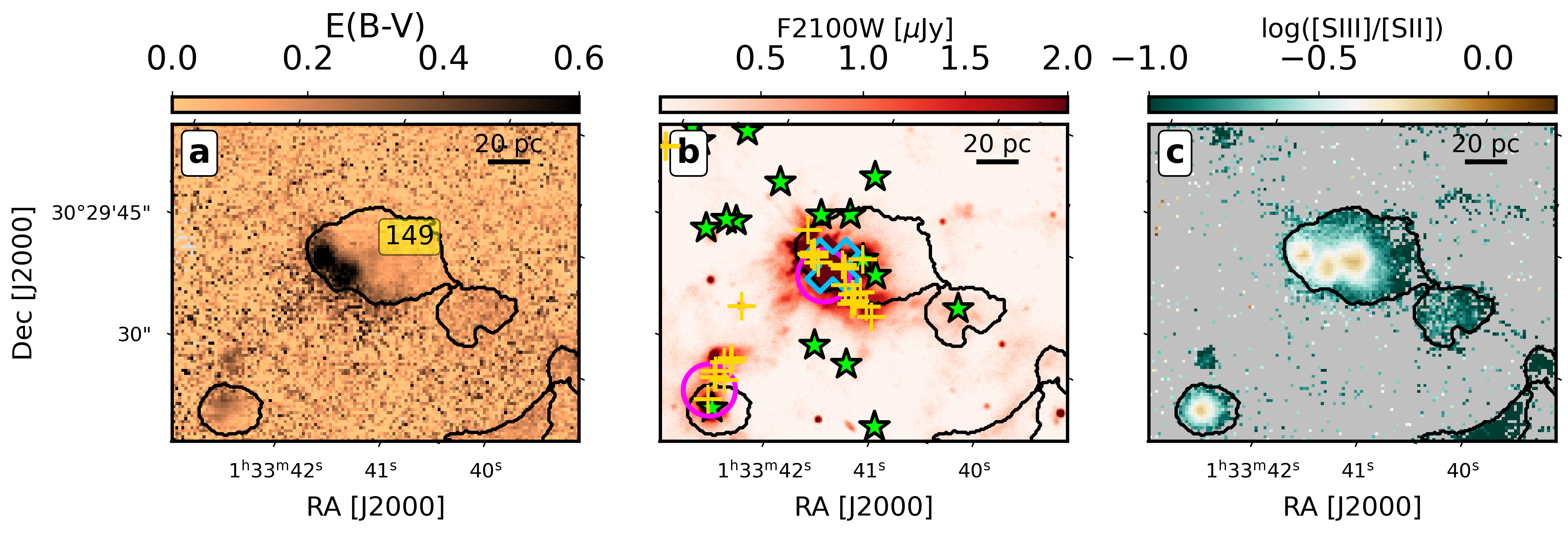}
  \caption{Zoom-in 2D maps of ID\,149. From left to right: E(B-V), MIRI F2100W, and \siiil/\siit\ maps. Green stars indicate massive stars (Sect. \ref{sec:phatter}), the large open cyan cross indicates a young stellar cluster, and magenta circles are giant molecular clouds from \citet[][Sect. \ref{app:ysc}]{Corbelli2017}. Yellow plus symbols refer to the YSOs from \citet[][Sect. \ref{sec:yso}]{Peltonen2024}. }
        \label{fig:zoom_ebv_137}
  \end{figure*}

    \begin{figure*}[!h]
\centering
\includegraphics[width=1.0\textwidth]{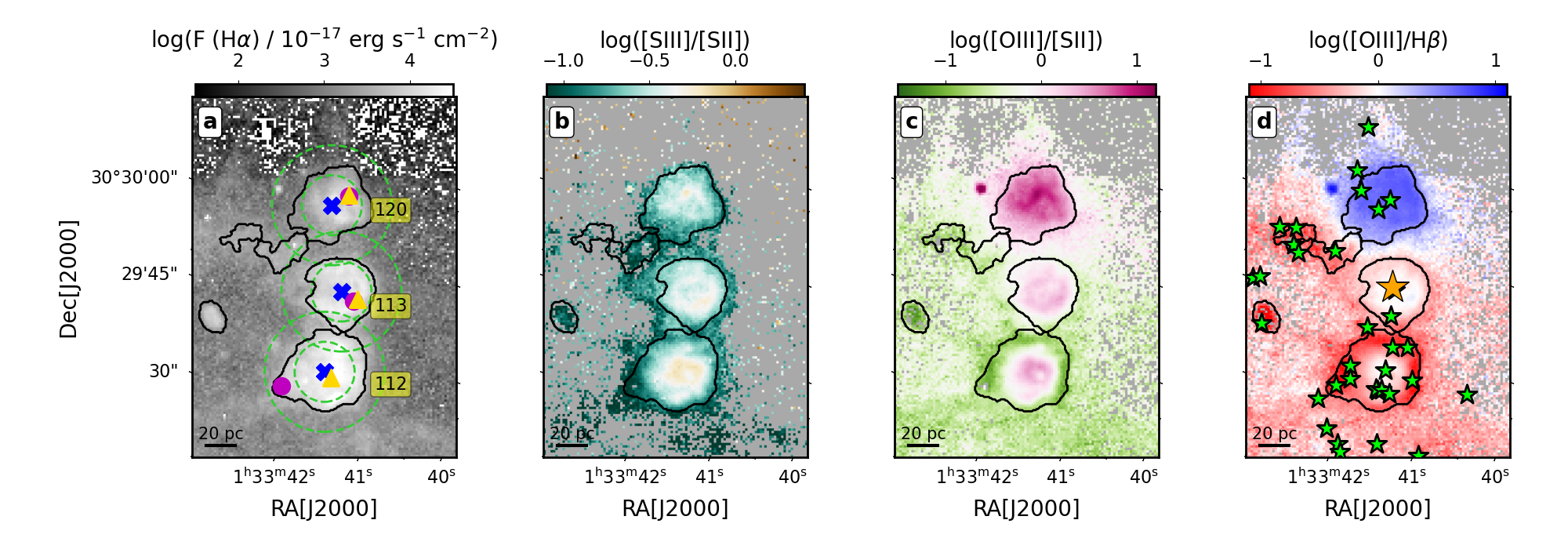}
  \caption{Zoom-in line-ratio maps of three example \hii\ regions (IDs\, 112, 113 and 120 from bottom to top). From left to right: Maps of \ha\ flux, \siii/\sii, \oiii/\sii, and \oiii/\hb\ ratios. 
 In panel a, the blue crosses are the geometrical centroids of the nebular mask, the magenta circles, and the green triangles indicate the peaks of the \ha\ and \oiii\ emission within the nebular mask, respectively. The dashed green apertures have a radius of 5 and 10\arcsec, respectively. In panel d, green and orange stars indicate massive stars (Sect. \ref{sec:phatter}) and WR stars (Sect. \ref{sec:wr}), respectively. }

        \label{fig:ion_fronts_zoom}
  \end{figure*}

\subsection{Dust attenuation and emission}\label{sec:ebv}

As an illustrative example, we discuss here \hii\ region ID\,149, which exhibits a wide range of E(B-V) values within it, as shown in Fig. \ref{fig:zoom_ebv_137}a. Specifically, there are two dusty clumps (E(B-V) $> 0.4$) at the left edge of the ID\, 149 nebular mask. This is one of the few areas strongly affected by dust reddening in our data.
While none of the ionising stars selected in Sect. \ref{sec:phatter} lie within these dusty clumps (indicated as green stars in Fig. \ref{fig:zoom_ebv_137}b), the \ha\ emission is intense (see Fig. 
\ref{fig:M33_muse}), indicative of on-going star-formation that could be embedded in dust. By looking at JWST data from \citet[][Sect. \ref{sec:yso}]{Peltonen2024}, we find several YSOs clustering around these two dusty knots (Fig. \ref{fig:zoom_ebv_137}b), acting as heating source of the gas within these clumps. This cluster of YSOs was already identified by \citet[][Sect. \ref{app:ysc}]{Corbelli2017} as a young stellar cluster (cyan open cross) associated with a molecular cloud (magenta circles).
Interestingly, the \siii/\sii\ map (zoomed-in view in Fig. \ref{fig:zoom_ebv_137}c) reveals three blobs of intense \siii/\sii\ emission within the \hii\ region ID\,149. Two of these correspond to the high-dust attenuation regions powered by YSOs described above, while the nebular emission closer to the centre of the nebular mask is likely powered by YSOs embedded in a lower dust content. In contrast to simple \hii\ regions, where the ionisation parameter exhibits a smooth radial gradient around the central ionising source, complex systems such as ID\,149, hosting multiple ionising sources, display a spatially non-uniform and locally modulated ionisation parameter distribution across the nebula.

This example highlights the effectiveness of the \siii/\sii\ ratio as a diagnostic tool for the ionisation parameter. In fact, ID\,149 is a complex emission regions, hosting multiple ionising sources, including massive stars and young stellar clusters. A comprehensive interpretation of these data requires advanced modelling approaches that account for the contributions of multiple ionised clouds, such as the HOMERUN photoionisation framework \citep{Marconi2024}.

\begin{figure}
\centering
\includegraphics[width=0.4\textwidth]{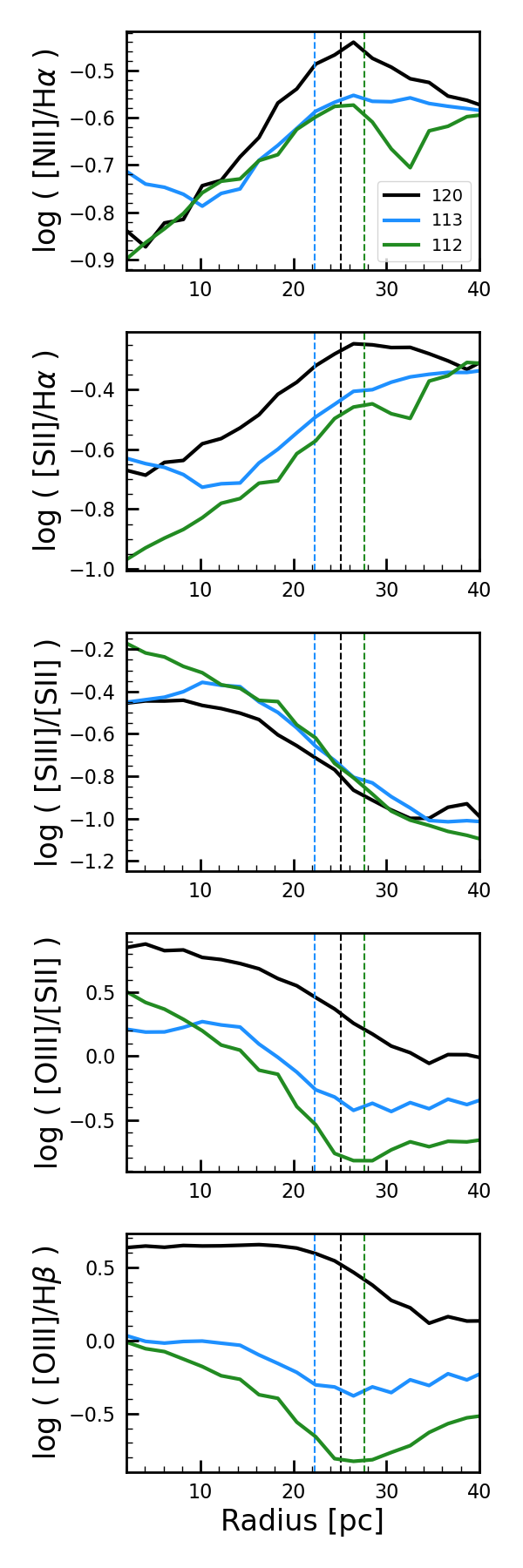}
  \caption{Radial profiles of the \nii/\ha, \sii/\ha, \siii/\sii, \oiii/\sii, and \oiii/\hb\ ratios. Three different \hii\ regions (IDs\,112, 113, 120) are represented by different colours as labelled in the legend. Dashed vertical lines represent the sizes of the nebular masks defined in Sect. \ref{sec:astrodendro}.}
    \label{fig:radial_profiles}
\end{figure}

\subsection{Resolving the \hii\ region ionisation fronts}\label{sec:neb_geo}

The \hii\ regions in our dataset feature different nebular structures and ionising sources, as showcased by the three example nebular regions numbered in Fig. \ref{fig:ion_fronts_zoom}a. The \hii\ region ID\,113 exhibits a projected nearly spherical geometry with emission from higher ionisation lines (i.e. \oiii) peaking in the centre and that from low-ionisation lines gradually increasing from the centre outwards. The \hii\ region ID\,120 is characterised by a smoother variation of line ratios within the nebular mask, with \oiii/\hb\ emission roughly constant throughout the entire region. As discussed later (Sects. \ref{sec:radial_tot} and \ref{sec:oiii_bubbles}), this may be suggestive of ionising photon leakage  (i.e. a density-bounded \hii\ region) in ID\,120, in contrast to ID\,113 which, with well-defined ionisation fronts at the edges of the nebular mask, is more representative of the nebular properties from fully ionised (i.e. ionisation-bounded) \hii\ regions. \hii\ region ID\,112 is another example of ionisation-bounded region and exhibits well-defined ionisation fronts within the nebular mask, with line ratio variations more pronounced than those of ID\,113.

ID\,113 hosts a WR star but does not exhibit an \oiii/\hb\ ratio as extreme as that of ID\,120, which harbours three ionising stars, none of which classified as WR. This finding highlights that intense and extended \oiii\ emission is not exclusive to WR-driven nebular emission, as already discussed in Sect. \ref{sec:hii_int}. 

The variety of nebular properties displayed by the \hii\ regions in our sample is reflected in the radial profiles of their line ratios. 
As illustrated in Fig. \ref{fig:ion_fronts_zoom}a, we considered concentric annuli with steps of 0.5$''$, starting from a radius of 0.5$''$ up to 12$''$, centred on the geometrical centroid of the dendrogram (blue crosses), defined as the mean of the coordinates of the nebular mask. We extracted the integrated spectrum from each annular ring (i.e. the region enclosed between each pair of adjacent annuli), masking the emission from other nebular regions. We chose to use the geometrical centre to avoid extracting spectra centred on peaks of \ha\ or \oiii\ emission (magenta circles and green triangles in Fig. \ref{fig:ion_fronts_zoom}a, respectively) which can be offset from the centre of the nebular mask. This is particularly evident in ID\,112 (left panel) where the \ha\ peak is clearly displaced from the geometrical centre of the mask.

Figure \ref{fig:radial_profiles} presents the radial profiles of the mean \nii/\ha, \sii/\ha, \siii/\sii, \oiii/\sii\ and \oiii/\hb\ ratios measured in concentric annuli, for the three \hii\ regions described in the previous section. Both \nii/\ha\ and \sii/\ha\ ratios increase outwards until \sii\ and \nii\ reach their peak emission at outer edge of the nebula \citep[see theoretical predictions from][]{Jin2022}, i.e. at the ionisation front. After this, we observe a plateau probably associated with the DIG.

The radius at which the ionisation front is reached corresponds rather accurately to the size of the nebular masks (vertical dashed lines), defined as the mean distance of the furthest 25\% pixels from the centre of the dendrogram.
This correspondence demonstrates the ability of our procedure in isolating physically justified ionisation regions.

The \siii/\sii\ (Fig. \ref{fig:ion_fronts_zoom}b and Fig. \ref{fig:radial_profiles}) decreases from the centre outward, mainly tracing the decline in ionisation parameter. The \oiii/\sii\ and \oiii/\hb\ radial profiles of ID\,120 differ from those of ID\,113 and ID\,112. Specifically, the \oiii/\sii\ (Fig. \ref{fig:ion_fronts_zoom}c and Fig. \ref{fig:radial_profiles}) keeps decreasing radially in ID\,120, while there is a plateau at the ionisation front of ID\,113 and 112. The \oiii/\hb\ ratio (Fig. \ref{fig:ion_fronts_zoom}d and Fig. \ref{fig:radial_profiles}) of ID\,120 remains constant within the first 20 pc (5\arcsec) before decreasing sharply, while \oiii/\hb\ decreases steadily from the centre outwards in ID\,113 and ID\,112 until the ionisation front is reached. 

Given the similar behaviour of the \siii/\sii\ ratio for the three \hii\ regions, a difference in ionisation parameter cannot account for the variations of radial profiles among \hii\ regions. The primary difference lies in the \oiii\ emission, which can be attributed to differences in the hardness of the ionising radiation or gas physical conditions and optical depth of the nebula \citep[e.g. ][]{Brinchmann2008}. The reasons for these variations in radial profiles are discussed in more details is Sects. \ref{sec:discussion}.

\section{Stacked radial profiles}\label{sec:radial_tot}

We investigated the radial profiles of line ratios by computing median profiles in the largest \hii\ regions of our sample. For this purpose, we first selected the nebular masks with more than 600 pixels and we intentionally excluded filamentary and asymmetric structures, as constructing radial profiles for such regions would not provide representative results due to their elongated geometries. Specifically, we refined our selection to the \hii\ regions with an aspect ratio (i.e. the ratio of the major to minor axis) of less than 1.5. This ensures the selection of 32 \hii\ regions with nearly spherical geometries (see Appendix \ref{app:AppG}). 

To classify the 32 selected \hii\ regions into three representative subsets and minimising the variance within each subset, we applied a K-means clustering algorithm \citep{Lloyd1982,Elkan2003} based on the logarithmic line ratios of \oiii/\hb\ and \oiii/\sii\ (see bottom panel of Fig. \ref{fig:AppG}).

We chose these line ratios because the first traces the ionisation level and the hardness of the ionising radiation, while the latter, in addition to its dependence on the ionisation parameter, has been proposed as a diagnostic of optical depth variations within \hii\ regions \citep{Pellegrini2012}. 

The `high-ionisation' subset exhibits the highest line-ratio values and comprises five \hii\ regions with both \oiii/\hb\ and \oiii/\sii\ $ >1.6$, three of which hosting WR stars (IDs\,40, 50 and 126), and the subset overall traces harder radiation fields. The `low ionisation' subset is characterised by lowest \oiii/\hb\ ($<0.4$) and \oiii/\sii\ (<0.3) ratios, likely reflects more evolved stellar populations dominated by older or less massive stars, and includes nine \hii\ regions. The remaining regions fall into an `intermediate' subset with line-ratio values between those of the other two subsets and which likely represent typical ionisation conditions.

\begin{figure}
\centering
\includegraphics[width=0.38\textwidth]{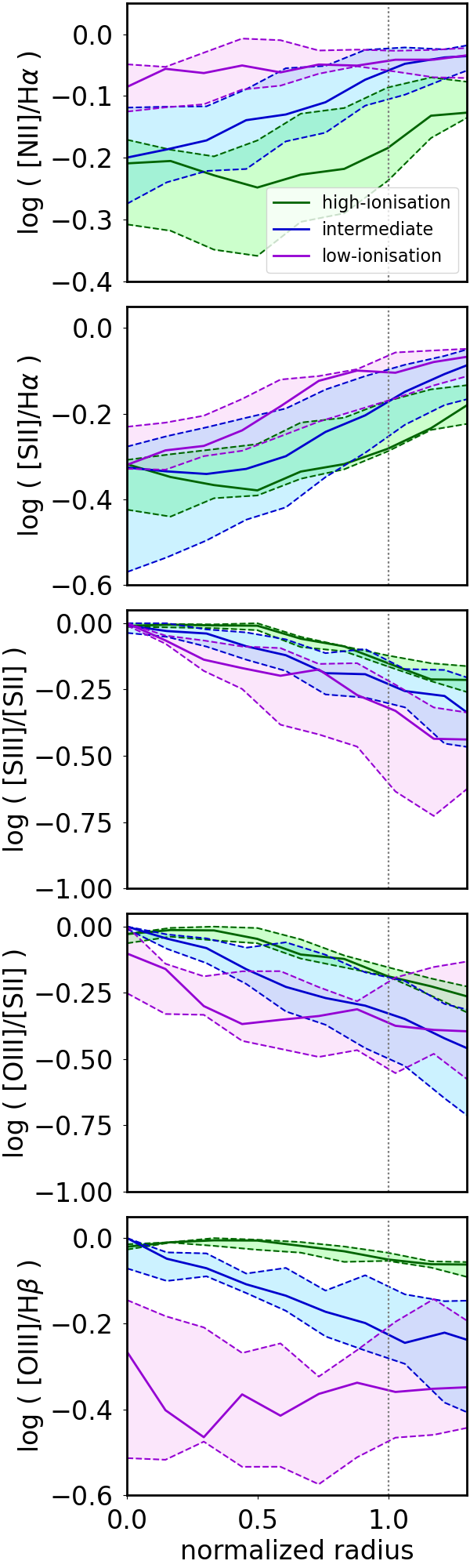}
  \caption{Median line-ratio profiles for three subsets of 32 \hii\ regions selected as described in Sect. \ref{sec:radial_tot}. The subsets are colour-coded as labelled in the legend.}%
        \label{fig:radial_tot}
\end{figure}

We computed the radial profiles from the spectral measurements extracted using concentric annuli, from 0.5\arcsec\ up to 12\arcsec, as described in Sect. \ref{sec:neb_geo}, for each of the 32 selected \hii\ regions. The radial profiles were normalised to the peak value of the line ratio of interest and the radius to the nebular mask size as defined in Sect. \ref{sec:neb_geo}. We then computed the median radial profile of the  \nii/\ha, \sii/\ha, \oiii/\sii\, \siii/\sii\ and \oiii/\hb\ ratios for each subset, which are shown in Fig. \ref{fig:radial_tot}.

The line-ratio radial profiles of the `high-ionisation' subset differ significantly from the others. Their \nii/\ha\ and \sii/\ha\ ratios continue to increase beyond the nebular boundary (which is the normalised value of the nebular mask size, vertical dotted line in Fig. \ref{fig:radial_tot}), whereas in the other subsets these ratios tend to flatten approximatively at these distances, indicating the presence of well-defined ionisation fronts. The \siii/\sii\ and \oiii/\sii\ of the `high-ionisation' \hii\ regions start decreasing at larger distances from the centre and the decline of $\approx$0.2 dex is more gradual compared to the steeper drop of ~0.4 dex observed in the other cases. Finally, the radial profile of \oiii/\hb\ of the `high-ionisation' subset is nearly constant ($\approx$ 0.1 dex decline) throughout the nebula. 
The rise of \oiii/\hb\ beyond the nebular bboundary in the `low-ionisation' subset is due to contamination from nebular emission outside the nebular mask (see top panel of Fig. \ref{fig:AppG}).

\section{Discussion}\label{sec:discussion}

\subsection{The different origins of intense \oiii\ emission} \label{sec:oiii_bubbles}
 
\begin{figure*}
\centering
\includegraphics[width=0.85\textwidth]{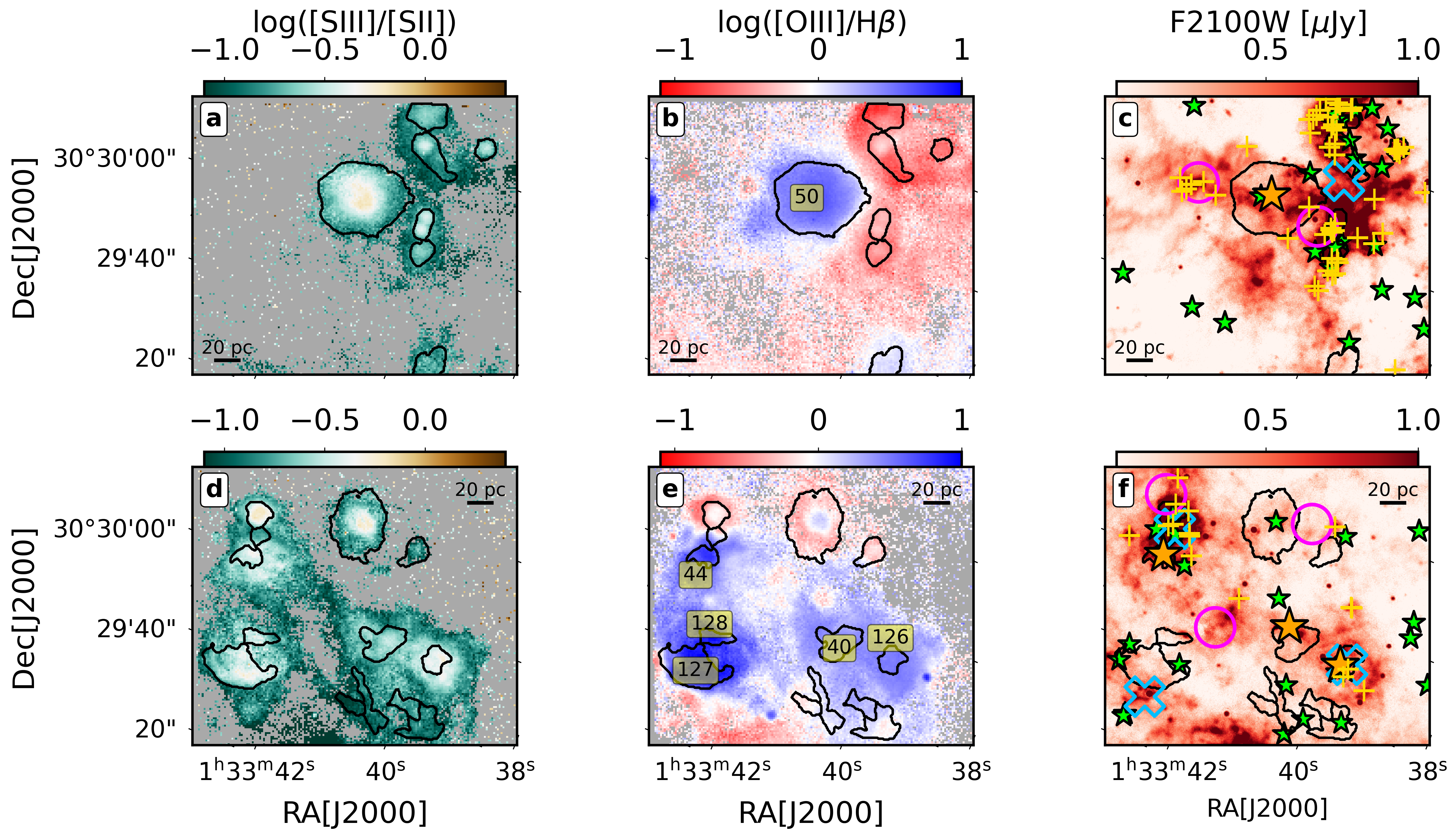}
  \caption{ Zoom-in 2D maps of some nebula with intense \oiii\ emission, namely ID\,40, ID\,44, ID\,50, ID\,126, ID\,127, and ID\,128. From left to right: \oiii/\hb, \siii/\sii, and MIRI F2100W maps. Green and orange stars indicate massive stars (Sect. \ref{sec:phatter}) and WR stars (Sect. \ref{sec:wr}), respectively. Yellow plus symbols are young sterllar clusters from \cite{Corbelli2017} and the magenta circles, when present, indicate the associated molecular clouds. Cyan empty crosses refer to the YSOs from \cite{Peltonen2024}. }
        \label{fig:oiii_bubbles}
  \end{figure*}

As noted in Sects. \ref{sec:obs_prop} and \ref{sec:zoom-in}, certain regions are characterised by strong \oiii\ emission irrespective of the presence of a WR star. ID\, 50 (Fig. \ref{fig:oiii_bubbles}, top panels) and ID\,113 (Fig. \ref{fig:ion_fronts_zoom}) are notable examples of \hii\ regions hosting a central WR star and exhibiting high \oiii/\hb\ ratios throughout all the nebular mask. The ionisation parameter, as traced by the \siii/\sii\ (Figs. \ref{fig:oiii_bubbles}a and \ref{fig:ion_fronts_zoom}b), peaks at the centre and decreases outwards.  Other \hii\ regions hosting WR stars, such as IDs\,40, 50 and 126 (bottom panels of Fig. \ref{fig:oiii_bubbles}), are characterised by intense \oiii\ emission which is embedded in a more extended \oiii\ halo, consistent with the results from \cite{Tuquet2024}. 

Conversely, ID\,120 (see Fig. \ref{fig:ion_fronts_zoom}) and ID\, 127 (Fig. \ref{fig:oiii_bubbles}, bottom panels), despite exhibiting extreme line ratio similar (or even higher in the \oiii/\hb) to the WR-driven nebulae, do not host any WR star reported in the literature. Both ID\,50 and 120 appear approximately spherical in projection, show a constant \oiii/\hb\ across their extent, and lack clear ionisation fronts. This is of particular interest because typical ionisation-bounded \hii\ regions exhibit ionisation stratification, with \oiii\ and \siii\ emission being stronger near the ionising source and weakening outward, where the emission from lower-ionisation species, such as \nii\ and \sii, becomes dominant. ID\,127 is even more intriguing as it is embedded within a more extended halo of \oiii\ emission (Fig. \ref{fig:oiii_bubbles}d) that also encompasses other regions. 

We discuss here the different factors that could drive the high \oiii/\hb\ ratio, which is often linked to a high ionisation parameter. In ID\,120, however, the \siii/\sii\ decreases from the centre outward, indicating a radial decline in ionisation parameter. This suggests a scenario where the hardness of the ionising radiation remains sustained, while the ionisation parameter declines outward. ID\,127 exhibits approximatively one order of magnitude variations of \siii/\sii\ within the nebula and a localised region of enhanced \siii/\sii\ values that do not coincide with the centre of the nebular mask, suggesting spatial variations in ionisation conditions. The \siii/\sii\ ratio in both regions is relatively low compared to other nearby regions, indicating a comparatively lower ionisation parameter.  

Another primary driver of intense \oiii\ is a hard ionising radiation field, such as that produced by hot, massive stars (e.g. O-type and WR stars). 
While the presence of WR stars can explain the emission observed in ID\,50 and ID\,113, the stellar ionising source driving the emission of ID\,120 and ID\,127 is likely of different origin. At this stage, it is not yet clear whether WR stars are absent in these regions or whether their signatures are simply undetectable with current data.
Unfortunately, our data lack spectral coverage of the \heii\ recombination line ($>54.4$ eV) preventing us to directly trace the hardness of the ionising radiation fields. However, the \hei\ emission is spatially resolved in six regions hosting WR stars, as well as in some hosting YSSCs and YSOs. The detection of \hei\ recombination emission (requiring ionising photons $>24.6$ eV to produce He$^{+}$), along with collisionally excited \oiii\ (ionisation potential of $35.1$ eV) observed in all the regions hosting WRs and even in ID\,120 suggests the presence of a hard ionisation field even in the latter. Interestingly, spatially resolved \hei\ emission is barely detected in ID\,127 making it even more challenging to attribute its intense \oiii\ emission solely to hard ionising radiation. We also checked several X-ray catalogs \citep{Grimm2005, Tuellmann2011,  Williams2015, Yang2022, Lazzarini2023} to search for potential sources of hard ionising radiation, such as ultra-luminous X-ray sources (ULXs), high-mass X-ray binaries, or shocked gas regions, but we did not find any spatial association with the regions of extended \oiii\ emission.

To further investigate the nature of the ionising source of the extreme \oiii\ emission, we extracted the spectra of the three stars (two selected as described in Sect. \ref{sec:phatter} and an additional one with F475W--F814A$=$0.775), lying close to the boundaries of ID\,127 nebular mask, as well as of the three stars within ID\,120. None of the spectra exhibit WR features and the detection of O-type stars features is hindered by the intense nebular emission, which challenges the extraction of uncontaminated spectra. In addition, these regions are devoid of YSOs as selected by \cite{Peltonen2024}. We note, however, that there is a YSCC not associated with any molecular cloud (class c3 of \citealt{Corbelli2017}, see Appendix \ref{app:ysc}) in the south-east of ID\,127. This is co-spatial with a very compact infrared source as visible from the MIRI images (Fig. \ref{fig:oiii_bubbles}f). While this suggests ongoing star formation, we cannot confirm whether it contributes significantly to the ionisation of the region.

Moreover, the nebular geometry has an important influence on collisionally excited lines, rather than Balmer recombination lines \citep{Kennicutt2012}. In a shell-like structure or hollow or filamentary density distribution, the ionising photons could be leaking out in some directions while still exciting O$^{2+}$ throughout the region. We also should not neglect projection effects. Indeed, the apparent spherical symmetry in projection might not reflect the true 3D structure. The \hii\ region could have a clumpy or filamentary nature, where ionising photons are leaking outward through channels of low optical depth along multiple sightlines. The regions ID\,120 and ID\,127 have a weaker IR emission (Fig. \ref{fig:dust_maps}b) relative to other regions, indicative of a lower dust content. While the role of dust in regulating the escape of ionising photons remains unclear, low optical depth regimes would favour photon leakage, as expected in environments with little or no dust \citep{Pellegrini2012}. This complex interplay between photon leakage, gas physical conditions and nebular geometry could be the explanations for the emission line properties of the nebula with intense \oiii\ emission and their surrounding gas.

\subsection{Perspectives from theoretical models}

Simulations of emission-line ratio mapping from \cite{Pellegrini2012}, combining the photoionisation code {\sc CLOUDY} \citep[version c08.00;][]{Ferland1998} and the 2D surface-brightness simulator {\sc SurfBright}, demonstrate that ionisation stratification within a nebula strongly depends on optical depth, allowing one to differentiate nebulae that are fully ionised (i.e. optically thick, ionisation-bounded) from those leaking ionising photons (i.e, optically thin, density-bounded) nebulae. Their low-optical-depth models do not show a low-ionisation transition layer, i.e. they lack  clear ionisation fronts, as exemplified by ID\,120 in our sample (Fig. \ref{fig:ion_fronts_zoom}). In contrast, the low-ionisation envelopes of optically thick nebulae resemble classical, spherical Str\"omgren spheres (e.g. ID\,46 or 113, Figs. \ref{fig:ion_fronts_zoom} and \ref{fig:oiii_bubbles}), that is, ionisation-bounded \hii\ regions, in agreement with the scenario described in the previous section (Sect. \ref{sec:oiii_bubbles}). 

Comparing photo-ionisation models of ionisation-bounded versus density-bounded \hii\ regions, \cite{Plat2019} carried out a thorough study of the impact of escape fraction of Lyman-continuum (LyC) photons on observed emission-line ratios. In general, increasing the leakage of ionising photons progressively removes the lower-ionisation layers of the \hii\ region, increasing the ratios of high-to-low ionisation collisionally excited lines. However, several works \citep{Jaskot2013, Nakajima2014, Stasinska2015, Plat2019}  highlight the observational challenge of tracing LyC photons leakage due to the competing effects and degeneracies of the ionising photon escape fraction with other physical parameters of the emitting nebulae, such as the ionisation parameter, gas density, metallicity, age of the stellar populations and the presence of multiple ionising stars. Therefore, we refrain from attempting the estimation of the LyC photon escape fraction from the observed line ratios.

\cite{Pellegrini2020} presented time-dependent predictions of line emission from \hii\ regions surrounding evolving star clusters, showing that line ratios vary over time  because of the evolution dependence of model parameters. For example, pressure evolution causes \hii\ regions in clouds of finite mass to shift between being density-bounded (i.e. lower optical depth) and  ionisation-bounded (i.e. higher optical depth to ionising radiation). Although the predicted line ratios can migrate across different regions of the BPT diagram, they can still be interpreted probabilistically. The analysis of \cite{Pellegrini2020} shows that the regions with high \oiii/\hb\ ratios tend to be associated with high star formation efficiency (SFE, i.e. fraction of gas mass converted into stars). A fraction of the spaxels of ID\,120 can reach low \nii/\ha\ and high \oiii/\hb\ values populating the area where density-bounded, high SFE, low density and optically thin models lie \citep[Fig. 9 of][]{Pellegrini2020}. The low \nii/\ha\ values at high \oiii/\hb\ observed in our data are consistent with models predicting high ionising photon escape fractions, as shown in \citet[][see their Fig. 9a]{Plat2019}.

To conclude, \cite{Jin2022} analyse how strong optical emission lines (i.e. \ha, \hb, \oiii, \nii, \sii\ and \oi) depend on the complexity of nebular geometry, finding that lines originating near the nebular boundaries are particularly sensitive to geometric complexity, whereas lines emitted throughout the nebula primarily reflect the internal ISM density distribution. This highlights the complex interplay between geometry and density structure in interpreting our observations.

\section{Summary and conclusions}\label{sec:conclusions}

We have presented, for the first time, observations of a 24 arcmin$^{2}$ (1.43 kpc$^2$) MUSE mosaic along the southern major axis of the local group galaxy M33, resolving its ionised nebulae with the unprecedented resolution of $\approx$5 pc. 

In this work, we performed a systematic identification of \hii\ regions powered by hot, massive stars and young stellar clusters, using the \texttt{astrodendro} Python package \citep{Robitaille2019}, combined with a subsequent watershed segmentation algorithm \citep{Vincent1991}. Based on the resulting nebular masks we extracted the integrated spectra of the 131 \hii\ regions in our sample and provide a catalogue of emission line measurements, along with a characterisation of their observed and physical properties, including \ha\ luminosity, dust attenuation and metallicity. 

We showcased a set of emission line-ratio maps and investigate the line-ratio radial profiles of \hii\ regions exhibiting different nebular properties. Our data revealed a remarkable diversity of \hii\ region nebulae in terms of their ionising stellar populations and nebular spatially resolved properties. We summarise the main results obtained from the analysis of these data below:

\begin{itemize}
\item[-] We identified \hii\ region nebulae down to \ha\ luminosities of $\approx 5\times$10$^{35}$ erg s$^{-1}$, one order of magnitude fainter than other surveys in local galaxies (Sect. \ref{sec:obs_prop}). On average these \hii\ regions have electron densities below 40 cm$^{-3}$ (i.e. the low density limit of the \siibl/\siirl\ ratio), dust reddening E(B-V) mostly below 0.3, with few more heavily attenuated regions (Sects. \ref{sec:hii_int} and \ref{sec:ebv}). The metallicity is slightly subsolar \citep[between 0.3 and 0.8 the solar value, assuming a solar 12+log(O/H) of 8.69;][]{Asplund2009} and we confirm a negative metallicity gradient of $\approx0.2$ dex within the first 2 kpc from the centre. 

\item[-] We showed that leveraging multi-band data is crucial to connect the energy sources driving feedback on the small scales (<10 pc) with the emission of the ionised gas. For example, multi-band catalogues of PNe and SNR enable us to clearly distinguish nebular emission driven by star formation (Sect. \ref{sec:class_hii}). Information on stellar populations (host massive stars, WR stars and young stellar clusters) are fundamental to unravel the sources of ionisation within \hii\ regions and interpret emission line properties (Sect. \ref{sec:spatial}). 

\item[-] We identified a variety of physical conditions across all \hii\ regions from spatially resolved flux and line-ratio maps (Figs. \ref{fig:line_ratio_maps} and \ref{fig:dust_maps}), including dust obscured regions powered by deeply embedded YSOs (Sect. \ref{sec:ebv}), regions characterised by different ionisation fronts, as traced by ratios such as \oiii/\sii\ (Sect. \ref{sec:zoom-in}), and highly ionised nebulae exhibiting enhanced \oiii/\hb\ emission (Sect. \ref{sec:oiii_bubbles}).

\item[-] Spatial mapping, along with line-ratio radial profiles (Sect. \ref{sec:radial_tot}), across regions with different \oiii/\sii\ and \oiii/\hb\ properties, indicate a mix of density-bounded (optically thin) and ionisation-bounded (optically thick) \hii\ regions. This interpretation aligns with the predictions from the `ionisation parameter mapping' technique proposed by \citet{Pellegrini2020}, which relies on spatial diagnostics involving the \sii/\oiii\ ratio and \ha\ emission as a probe of the optical depth of \hii\ regions.

\item[-] Our analysis further supports the effectiveness of using tools such as \texttt{astrodendro} to isolate emission associated with \hii\ regions. This approach enables for a more accurate delineation of the \hii\ region boundaries (Sect. \ref{sec:neb_geo}), a critical aspect for properly accounting for DIG contamination in emission-line analyses.

\end{itemize}

The presence of multiple ionising sources within individual nebular masks, combined with the complex interplay between their stellar properties and the spatial variations of physical conditions (e.g. density and ionisation parameter) within the gas, introduces significant complexity in interpreting observed emission features.  In particular, this calls for multi-cloud photoionisation modelling approaches, such as the HOMERUN framework \citep{Marconi2024}.

Modelling the complexity shown by our data requires advanced approaches that couple hydrodynamical simulations, capable of tracking the gas dynamics, shocks, and density variations, with detailed photoionisation codes that compute the emergent emission-line spectra. Such frameworks, as developed by \citet{Pellegrini2020} and \citet{Jin2022}, allow us to move beyond simplified static or single-source models, providing insights into the spatial and spectral signatures of feedback processes.
Similarly, high-resolution (1-4 pc) simulations such as those presented in \citet{Rathjen2023, Rathjen2025} offer valuable predictions on the evolution of ionised regions at parsec and sub-parsec scales. Implementing these state-of-the-art models is critical for fully exploiting the rich spatial and spectral information contained in our dataset. 

In summary, our analysis reinforces the idea that no two \hii\ regions are alike, especially at the small scales probed by our MUSE observations. This diversity reflects the intricate interplay between ionising sources, gas geometry, and local physical conditions, highlighting the need for flexible modelling strategies and high spatial resolution to fully capture nebular emission and stellar feedback processes in galaxies. Resolving the ISM at spatial scales of a few parsecs is therefore instrumental to directly probe the interaction zones between the ionising sources and their immediate environments, enabling the study of feedback processes and ionisation structures at their intrinsic physical scales, which would be otherwise diluted or unresolved in lower-resolution observations.

\section*{Data availability}

The catalogue data table \ref{tbl:cat} described in Sect. \ref{app:AppC} in only available in electronic form at the CDS via \url{http://cdsweb.u-strasbg.fr/cgi-bin/qcat?J/A+A/}. 
The reduced MUSE mosaic datacubes and maps will be made publicly available through the ESO Science Archive via the Phase 3 process.

\begin{acknowledgements}

Based on observations obtained at the ESO/VLT Programme ID 109.22XS.001 (PI: Cresci). AF acknowledges the support from INAF MiniGrant 2024 `The pc-scale view of \hii\ regions in M33' and from the project `VLT-MOONS' CRAM 1.05.03.07. AF, FB, FM \& GC acknowledge support from INAF Large Grant 2022 `The metal circle: a new sharp view of the baryon cycle up to Cosmic Dawn with the latest generation IFU facilities'. AF, FB, FM \& FM  acknowledge support from INAF Large Grant 2022 `Dual and binary SMBH in the multi-messenger era'.
IL, FM and AM acknowledge support from PRIN-MUR project `PROMETEUS'  202223XDPZM financed by the European Union -  Next Generation EU, Mission 4 Component 1 CUP B53D23004750006 and C53D2300080-006. GV acknowledges support from European Union’s HE ERC Starting Grant No. 101040227 - WINGS. FM acknowledges support from the INAF large grants 2023 `The Quest for dual and binary massive black holes in the gravitational wave era' and INAF large grants 2022 `The MOONS Extragalactic Survey'. This research made use of \texttt{astrodendro}, a Python package to compute dendrograms of Astronomical data (http://www.dendrograms.org/).
\end{acknowledgements}


\bibliographystyle{aa} 
\bibliography{M33biblio} 

\begin{appendix}

\section{Observing log}\label{app:AppA}
\begin{table}[!h]
\caption{Description of the MUSE observations for M33.}
\centering
\begin{tabular}{cccccc}
\centering
Pointing & RA & DEC & Obs. date  & PSF  \\
& hms & dms &  & \arcsec \\
\hline
1-3 & 01:33:44.273 & 30:33:54.770 & 06/09/22  & 0.5 \\
4-6 & 01:33:45.956 & 30:34:48.550 & 19/09/22  & 1.0 \\
7-9 & 01:33:47.638 & 30:35:42.327 & 20/09/22 & 1.5 \\
10-12 &  01:33:51.005 & 30:37:29.879 & 04/09/22 & 0.7 \\
13-15 &  01:33:42.591 & 30:33:00.999 & 25/09/22  & 0.7 \\
16-18 &  01:33:40.909 & 30:32:07.221 & 29/09/22  & 0.9 \\
19-21 &  01:33:39.230 & 30:31:13.444 & 03/10/22  & 0.8 \\
22-24 &  01:33:37.547 & 30:30:19.667 & 05/10/22  & 1.0 \\
\hline
\end{tabular}
\tablefoot{Each row corresponds to one OB observed on the same night within a 1 hour interval. The table includes pointing numbers corresponding to the OB (column 1, see Fig. \ref{fig:footprint}), pointing centre of the middle pointing of the strip (columns 2 and 3), data of observation (column 4), estimate of the seeing from the observatory DIMM monitor (column 5).} 
	  \label{table:obs} 
\end{table}

\section{Ancillary data} \label{app:AppB}

\subsection{PHATTER stellar photometry}\label{sec:phatter}

Our MUSE data largely overlaps with the area of M33 imaged by the PHATTER survey \citep{Williams2021}. PHATTER observed M33 with the ACS, WFC3-UVIS, and WFC3-IR on board HST obtaining near-UV (F275W, F336W), optical (F475W, F814W) and near-IR (F160W) imaging of a large contiguous area of M33 out to a radial distance of $\sim$ 3.5 kpc. Photometric catalogues have been derived using the DOLPHOT software package \citep{Dolphin2000} as described in \cite{Williams2021}.

\begin{figure}
\includegraphics[width=0.5\textwidth]{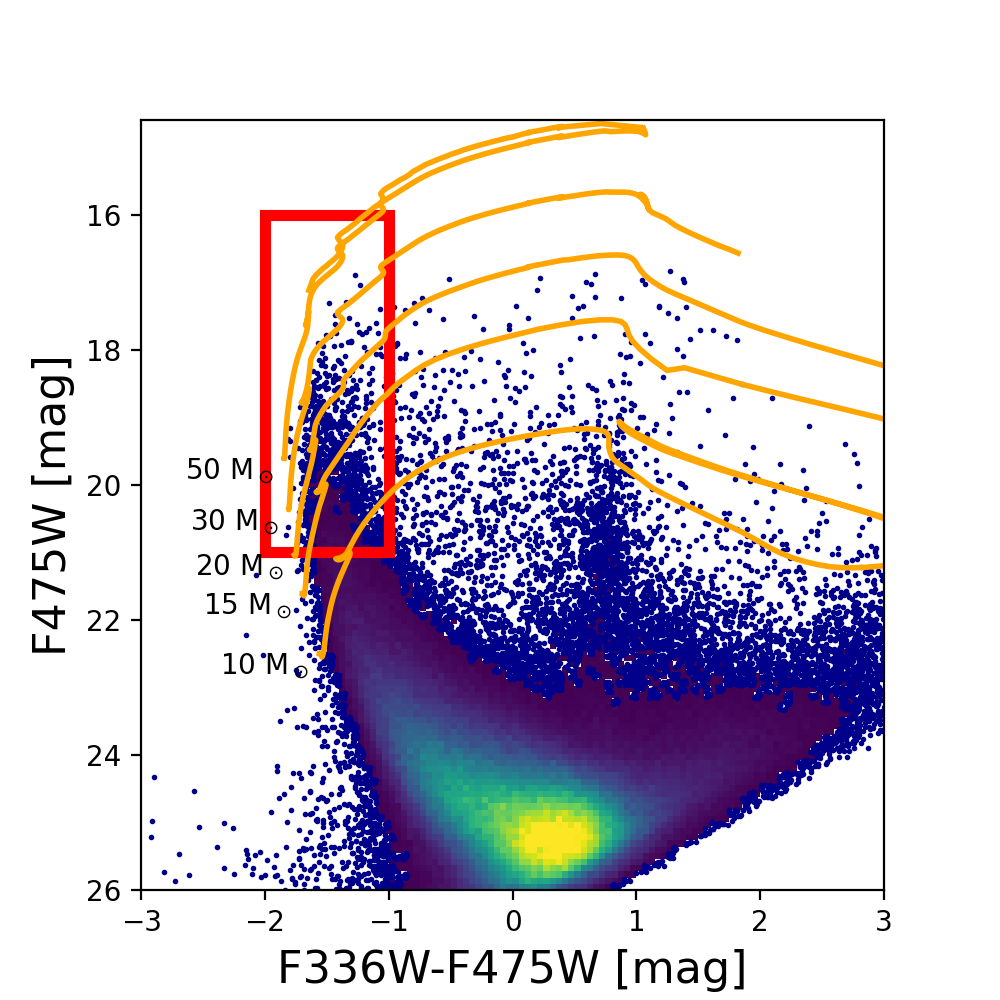}
\caption{Colour-magnitude diagram defined by the
F336W and F475W filters for the stars in the PHATTER catalogue\citep{Williams2021}. The red rectangle denotes the selection cuts for young main sequence
stars: F475W $< 21$ and $-2 <$ F336W$-$F475W $< -1$. Orange curves are evolutionary tracks \citep{Choi2016} along the main sequence and the red giant branch for stars of masses in the range 10 to 50 $M_{\odot}$, as labelled in the figure.}
    \label{fig:cmd}
\end{figure}

We use the PHATTER catalogue to select young main sequence stars  according to the following selection cuts: 
F475W $< 21$ and $-2 <$ F336W $-$ F475W $< -1$. 
These cuts select 662 stars within our MUSE footprint (Fig \ref{fig:M33_muse}). Fig. \ref{fig:cmd} shows our selection box in red on the colour-magnitude diagram defined by the F336W and F475W filters. Also shown the position of evolutionary tracks along the main sequence for stars of masses [10, 15, 20, 30, 50] $M_\odot$ computed from MIST v1.2 \citep{Choi2016} with metallicity $\rm [Fe/H] =-0.1$ at the distance of M33 (840 kpc) and foreground Milky Way reddening of $E(B-V) = 0.035$ mag.  

Our selection of F475W$=21$ corresponds to a zero-age main sequence 20 $M_\odot$ star with absolute magnitude of $M_{F475W} = -3.7$, or spectral type O6 according to the calibration of \cite{Castro2021}. While the HST data would allow us to extend the selection to fainter magnitudes and therefore later-type O stars, we do not do so here for two reasons. First, the current selection identifies the (main) ionising sources for a vast majority of identified \hii\ regions (albeit with some notable exceptions). Secondly, spectroscopic study of these stellar sources, which will be presented in a future work in this series, is limited to sources with F475W $\lesssim 19.7$. 

\subsection{Wolf-Rayet stellar catalogues}\label{sec:wr}
The WR stars are evolved massive stars which have lost their outer envelopes and shine as exposed He-burning stellar cores. They are characterised by notable broad emission lines of He, N, and C, and play an important role in the ionisation of the local ISM because of their hard ionising spectrum. 

Extensive searches using photometric selection and spectroscopic confirmation have led to a highly complete catalogue of WR stars in M33 \citep{Neugent2011,Neugent2019}. We obtain the positions of WR stars from the \cite{Neugent2011} catalogue, which is estimated to be complete to the 5\% level. Within the MUSE footprint we find 13 WR stars, 12 of which we spectroscopically confirm. Further details of the analysis of the WR spectra from the MUSE data are given in Appendix \ref{app:AppC}.

\subsubsection{Identification of WR stars in the MUSE data}
\label{app:AppC}

\begin{figure*}
\centering
\includegraphics[width=0.9\textwidth]{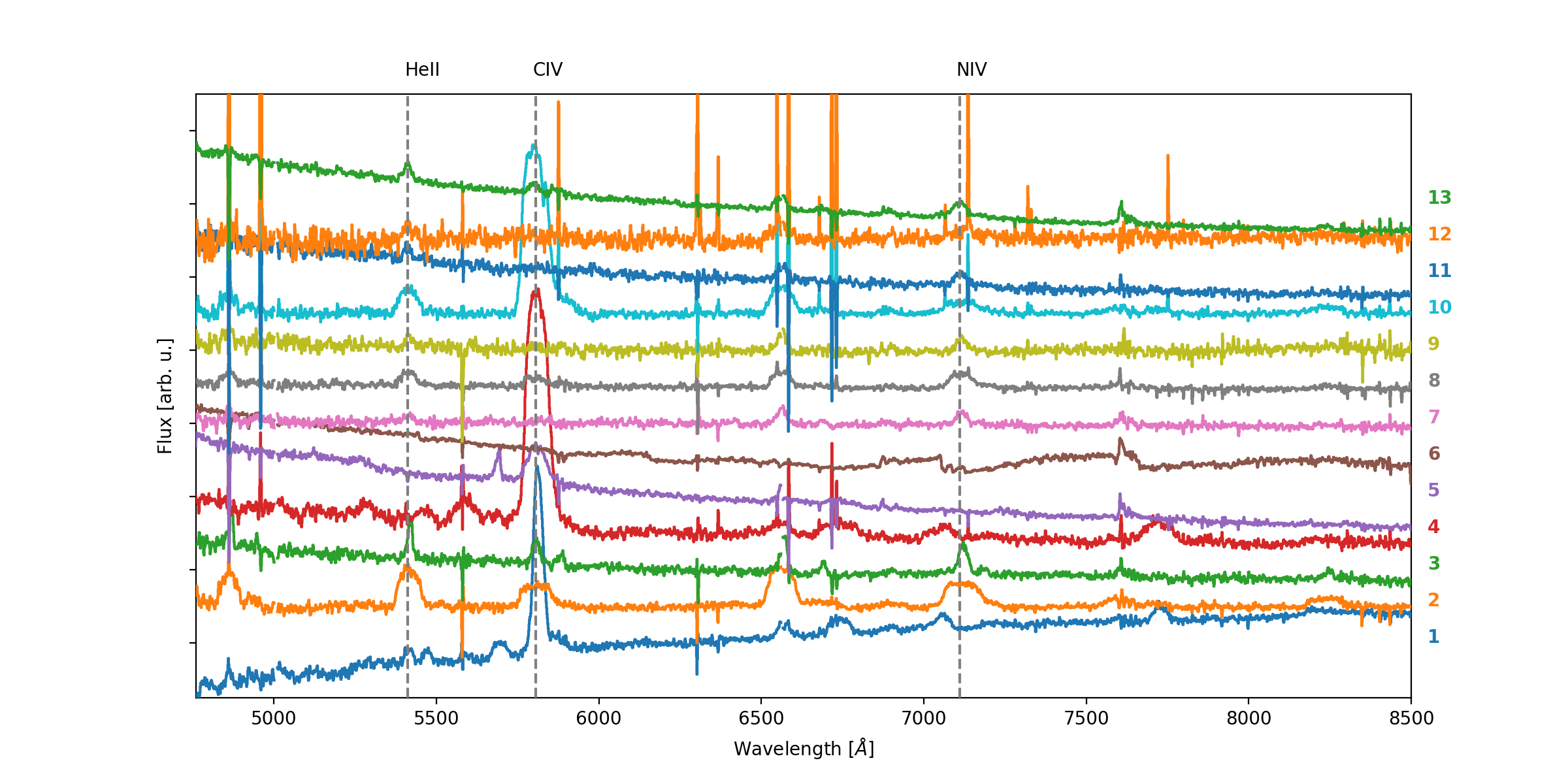}
  \caption{MUSE spectra of the WR stars from the \cite{Neugent2011} within our footprint. Colour-code refers to the identification ID in Table \ref{tbl:wr}. Flux is arbitrary for clarity. Vertical dashed lines indicate spectral signatures typically identified in WR spectra. The star ID6 is rejected as a WR candidate based on its spectrum.}
        \label{fig:AppC}
  \end{figure*}

By cross-matching the WR catalogue from \cite{Neugent2011} and the HST catalogue from \cite{Williams2021}, which we take as the astrometric reference, we find an astrometric offset of $\Delta(ra) = 0.614$ and $\Delta(dec) =0.207 $ which we correct. 

We extract the spectra of WR stars from the MUSE cube using a fixed aperture size of radius of 5 pixels ($1''$) and use an annulus of inner and outer radii of 9 and 11 pixels respectively ($1.8''$ and $2.2''$) for background subtraction. We exclude from the background aperture any pixels that overlap with other point sources detected in a MUSE white-light image. Nonetheless, because of extreme crowding, sometimes sources are oversubstracted.

Information on the WR stars within the MUSE footprint are in Table \ref{tbl:wr}, and their MUSE spectra are shown in Fig. \ref{fig:AppC}. We reject the star ID6 based on its MUSE spectrum.

\begin{table}
\caption{Wolf-Rayet stars from \cite{Neugent2011} within the MUSE footprint.}
\begin{tabular}{ccccc}
ID& Star ID NM11 & MJ98 & $m_V$ & Type\\
 &  &  & $\mathrm{mag}$ & \\
\hline
1 & J013338.20+303112.8 & WR55 & 20.65 & WC6 \\
2 & J013339.95+303138.5 & WR59 & 19.67 & WN4b \\
3 & J013340.04+303121.3 & WR60 & 19.76 & WNE$+$abs \\
4 & J013340.19+303134.5 & WR62 & 19.90 & WC4 \\
5 & J013342.53+303314.7 & WR71 & 18.99 &WC4-5 \\
6$^{*}$ & J013343.34+303534.1 & WR75 & 17.48 & --- \\
7 & J013345.58+303451.9 & WR79 & 20.75 & WNL \\
8 & J013345.99+303602.7 & WR80 & 20.43 & WN4b \\
9 & J013346.20+303436.5 & WR81 & 21.24 & WNE \\
10 & J013346.80+303334.5 & WR83 & 20.06 & WN6/C4 \\
11 & J013347.83+303338.1 & WR86 & 20.01 & WN4 \\
12 & J013350.23+303342.4 & WR91 & 21.14 & WN8-9 \\
13 & J013353.80+303528.7 & WR104 & 19.41 & WNE \\
\hline

\end{tabular}
\tablefoot{The first column identifies the spectrum in Fig. \ref{fig:AppC}, the second column gives the RA and DEC of the star in the \cite{Neugent2011} catalog. The third column identifies the star in the \cite{Massey1998} catalog, while the fifth column gives the apparent V-band magnitude from the same catalog. The sixth column reports the WR type. The star ID6 is rejected as a WR candidate based on its spectrum (see Fig. \ref{fig:AppC}). } \label{tbl:wr}
\end{table}

\subsection{Young stellar cluster candidates}\label{app:ysc}

\begin{figure}
\centering
\includegraphics[width=0.4\textwidth]{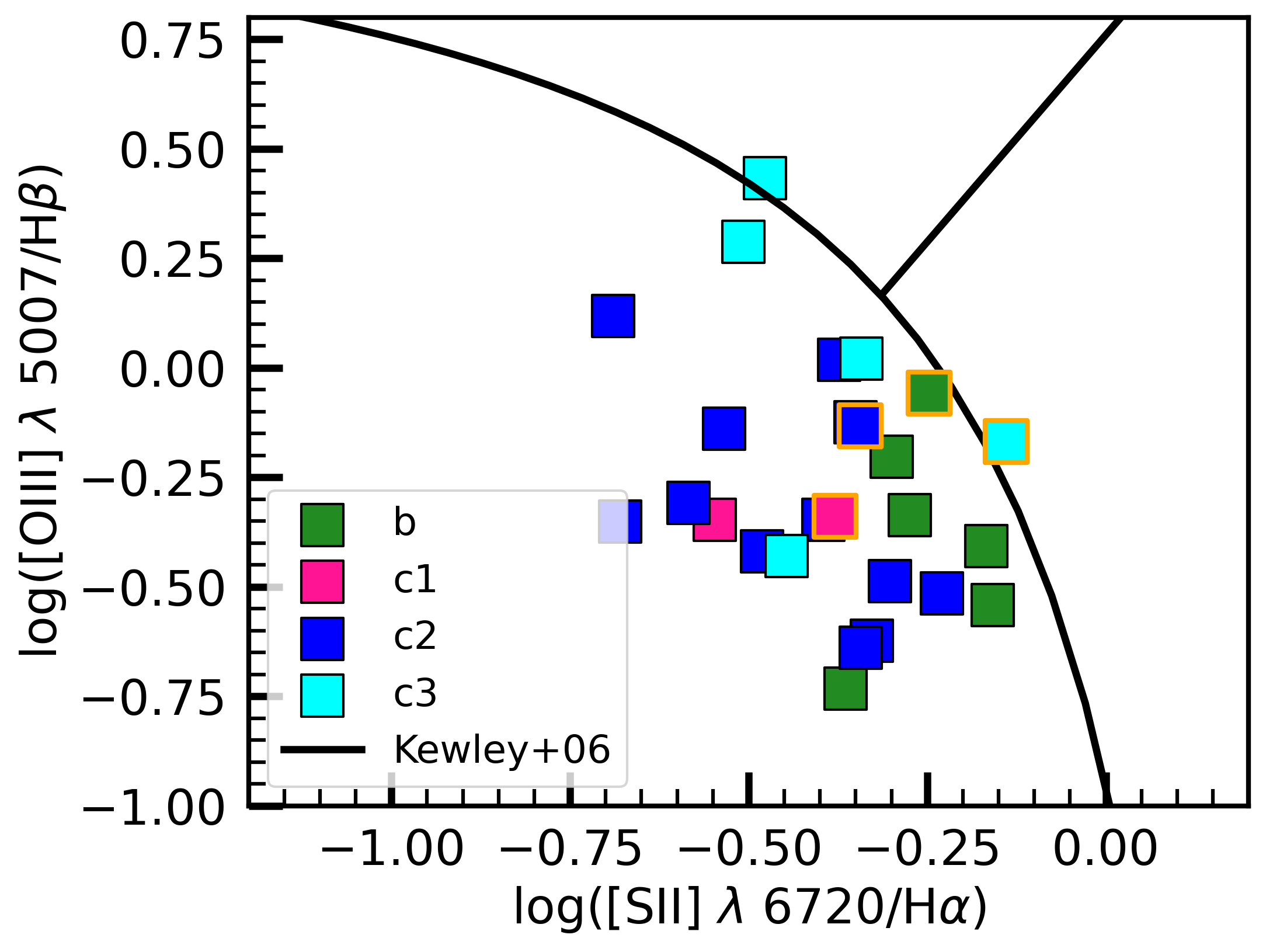}
  \caption{Type b and c YSCCs from \cite{Corbelli2017} on the \oiii/\hb\ versus \nii/\ha\ line-ratio diagram. The YSCCs types are colour-coded as labelled in the legend. The YSCCs with orange edges are those in the vicinity or co-spatial with SNRs. Continuous curve shows the demarcation criteria between the star-forming and active galactic nuclei by \cite{Kewley2006}.} \label{fig:ysc_bpt}
\end{figure}

Using multi-band data of M33 such as the CO(J=2-1) IRAM map \citep{Druard2014}, Spitzer-24 $\mu$m data \citep{Sharma2011}, GALEX-FUV, and H$\alpha$ maps,  \cite{Corbelli2017} investigated the association between giant molecular clouds (GMCs) and YSCCs. 

Specifically, \cite{Corbelli2017} identified 630 YSCCs across the whole star-forming disc of M33 using Spitzer-24 $\mu$m data. These YSCCs have been classified based on their observed properties in four classes: b, c, d and e. 
In particular, in this work we consider only YSCCs of class b and c (with H$\alpha$ or/and molecular cloud counterparts) as class d includes ambiguous sources, while class e YSCCs are contaminated by background sources, being visible only at 24$\mu$m and having no \ha\ or GMC counterpart. Accounting for the different classes of the YSCCs of \cite{Corbelli2017} in comparison samples \citep[e.g.][]{Johnson2022, Pflamm-Altenburg2013} is important to ensure a reliable interpretation of the results.

Class b YSSCs are in the embedded phase because they are associated with a GMC but lack an optical counterpart as \ha\ or FUV emission. Class c YSSCs have \ha\ counterparts but while the subclasses c1 and c2 are associated with GMCs, class c3 does not have an associated GMC. Class c1 and c2 differ because the first lack FUV detection. The sequence b-c1-c2-c3 has been interpreted as a time evolution of a collapsing GMC forming a young stellar cluster which evolves removing the native gas by stellar winds of massive stars \citep{Corbelli2017}.
Within the MUSE footprint, we find 30 type b and c YSCCs, of which 24 are also within the JWST 21 $\mu$m image. Only four YSCs from the \cite{Johnson2022} catalogue from the PHATTER project are common to the \cite{Corbelli2017} YSCCs considered here. This small overlap likely arises because of the age difference of objects in the two catalogs. Optical cluster catalogues have mostly stellar cluster older than 10 Myr, which are less likely to have H$\alpha$ counterparts, opposite to infrared selected YSCs.

We extracted the 1D-spectra within a circular aperture with a radius equal to 1.5 times the YSCC size (as defined by the 24 $\mu$m radius), and measured the emission line fluxes following the procedure described in Sect.~\ref{sec:linemaps}. Class b sources are located in regions where we did not identify any nebular emission mask and tend to occupy the extreme right side of the [SII]-BPT diagram, exhibiting a \sii/\ha\ $>$ 0.4 (Fig. \ref{fig:ysc_bpt}). We interpret this as clear indication that the observed emission originates primarily from the DIG component. This interpretation is consistent with expectations, given that class b YSCCs lack clear \ha\ counterparts, as noted in \citet{Corbelli2017}. On average, type-c3 YSCCs exhibit higher \oiii/\hb\ and  \sii/\ha\ ratios, likely due to reduced absorption of ionising photons, consistent with the lack of surrounding molecular material. Several YSCCs are found in proximity to nebular regions classified as SNRs, outlined in orange in Fig~\ref{fig:ysc_bpt}. This spatial coincidence may reflect the role of supernova-driven gas compression in triggering localised star formation, as shock fronts can enhance gas densities and promote gravitational collapse \citep[see also][]{Wainer2025}.

\subsection{Young stellar objects}\label{sec:yso}
Young stellar objects (YSOs) are stars in the early stages of formation and directly trace the onset of star formation.
\cite{Peltonen2024} presented 5 $\mu$m and 21 $\mu$m images of M33 obtained under the GO program 2128 and enabling, for the first time, the detection of YSOs beyond the Milky and the Magellanic Clouds. Combining these JWST mid-IR observations with HST images, \cite{Peltonen2024} identify 793 candidate YSOs in their footprint. These have been selected based on colours, visual inspection and proximity of GMCs (defined using CO(J=1-2) data from ALMA Atacama Compact Array). We find 336 YSOs candidates from this sample within our MUSE footprint. We find a strong spatial correspondence between thesse YSOs and the type-c YSCCs presented in the previous section (Sect. \ref{app:ysc}).

\subsection{Supernova remnants}\label{sec:snr}
\cite{Lee2014} produced a catalogue of SNR in M33 based on the \ha\ and \sii\ narrow-band images from the Local Group Galaxy Survey \citep{Massey2006, Massey2007}. They select remnants considering three criteria: \sii/\ha\ $>$ 0.4, a round or shell structure, and the absence of blue stars in the middle of the region. Sizes were estimated based on the extent of the region with \sii/\ha\ $>$ 0.4, considering the visible portion of the shell. In addition, we have checked that all the SNR candidates from \cite{Long2010} that fall in our M33 pointings have been selected. 

\subsection{Planetary nebulae}\label{sec:pne}

\cite{Ciardullo2004} performed a photometric and spectroscopic survey for planetary nebulae (PNe) across M33. They produced a catalogue of 152 PNe primarily identified through their prominent \oiiirl\ and \ha\ emission lines. We have verified that 8 PNe from \cite{Ciardullo2004} fall within our MUSE observations. We have also cross-matched with the PNe identified in \cite{Magrini2000, Magrini2001} and verified that they do not provide additional objects in our footprint.

\section{Nebular catalogue content}

Along with this paper, we release the catalogue of the 131 \hii\ regions identified in our MUSE footprint. The catalogue includes the \hii\ regions identification and coordinates, properties of the nebular mask (Sect. \ref{sec:astrodendro}), line flux measurements (Sect. \ref{sec:obs_prop}) and physical quantities (Sect. \ref{sec:hii_int}).  The exact content of the catalogue is described in Table \ref{tbl:cat}.

\begin{table*}[htp]
  \caption{Column description of the M$^{3}$D \hii\ region catalogue.}
\begin{center}
\begin{tabular}{cll}
\hline \hline
No. & Title & Description \\
\hline
  1 & HII\_ID    &  Nebular mask identification number\\
  2 & RA          & R.A. (J2000) in units of decimal degrees\\
  3 & DEC         & Decl. (J2000) in units of decimal degrees \\
  4 & Npix      &  Number of pixel within the nebular mask \\
  5 & star   &   Equal to `WR' when a WR star is within the nebular mask\\
  6 & YSCC &  Identification number of associated YSCC from \cite{Corbelli2017}\\
  7 & EBV   & E(B-V) value computed from the Balmer decrement \\
  8 & ne    & Electron density (in cm$^{-3}$) computed from the \siibl/\siirl\ doublet ratio \\
  9 & OH\_Scal   & Oxygen abundance in terms of 12+log(O/H) from the S-calibration of \cite{Pilyugin2016} \\
  10 & OH\_N2S2Ha & Oxygen abundance in terms of 12+log(O/H) from the N$_2$S$_2$H$\alpha$ indicator from \cite{Brazzini2024}\\
11-31 & [LINE]\_FLUX  & Line fluxes ($\rm 10^{-20} \, erg \, s^{-1} cm^{-2}$) summed over the entire nebular mask (see Sect. \ref{sec:astrodendro}) \\
& & corrected by dust attenuation \\
32-51   & [LINE]\_FLUX\_ERR & Errors on the line fluxes \\

\hline

\hline
\end{tabular}
\end{center}
\label{tbl:cat}
\end{table*}%

\label{app:AppD}

\section{Additional weaker emission lines and gas diagnostics}
\label{app:AppE}

\begin{figure*}[!h]
\centering
\includegraphics[width=0.9\textwidth]{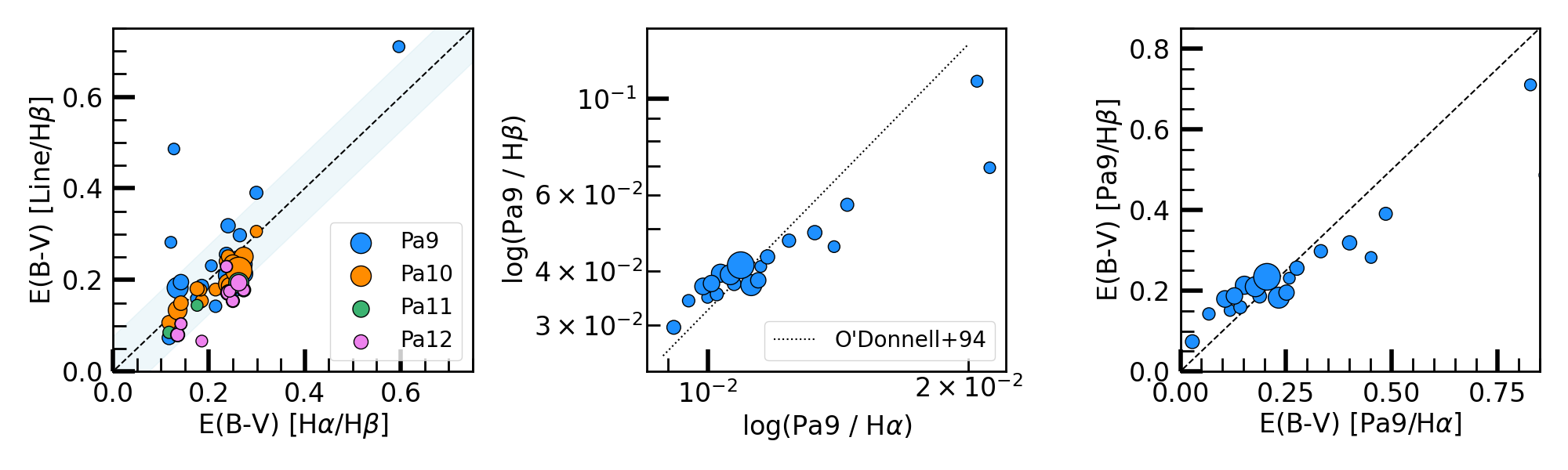}
  \caption{Comparison of dust reddening inferred using Balmer and Paschen lines. Left panel: E(B-V) computed assuming case B hydrogen recombination and using lines of the Paschen series (Pa$9$, Pa$10$, Pa$11$ and Pa$12$ as colour coded in the legend) versus E(B-V) from the Balmer decrement. Central panel: \paninel/\hb/ versus \paninel/\ha\ line ratios. Dotted gray line is the \cite{Odonnell1994} attenuation curve. Right panel: Comparison of E(B-V) inferred from the dust attenuation corrected \paninel/\hb\ and \paninel/\ha\ ratios. Dashed black lines indicate the 1:1 relations. Symbol sizes increases with increasing S/N, starting from S/N =5 up to 16.}
        \label{fig:attenuation}
  \end{figure*} 

Auroral lines are generally weaker compared to other nebular forbidden lines because they are highly temperature sensitive, making them valuable diagnostics for determining the metallicity of the nebula \citep[e.g.][for a recent review]{Kewley2019}. In our \hii\ regions, we find clear detections (SNR $>5$) of the auroral \niiau\ and \oiiau\ lines in 20 and 7 integrated spectra, respectively.

In a fraction of the spectra we also detect Paschen recombination lines (\paninel, \patenl, \paelevenl\ and \patwelvel) and \hei\ lines at 5016\AA, 5875\AA, 6678\AA\ and 7065\AA. Quantitatively, we detected (S/N $>5$) all four Paschen and \hei\ recombination lines in the spectra of 8 and 37 \hii\ regions, respectively. The single Pa$9$ line, strongest among the Paschen lines covered by the MUSE spectral range, has S/N>5 for 20 \hii\ regions. \hei\,5875\AA\  and \hei\,6678\AA\ are detected simultaneously with a S/N>5 in the spectra of 85 \hii\ regions. \\

We explore the possibility of using the \hei\ line ratios to constrain the density. Specifically, \texttt{pyNeb} predictions indicate that the \heid/\heib\ ratio is sensitive to electron density, though it exhibits a non-negligible dependence on electron temperature. Assuming an electron temperature T$_{\rm e}$ of 10$^4$ K, we find results consistent with the electron densities derived from the \sii\ doublet. Among the 38 \hii\ regions where both \heid\ and \heib\ are detected with a SNR$>5$, 34 regions show values approaching the low-density limit for \heid/\heib\ (i.e. $n_{\rm e}$ $<=50$ cm$^{-3}$), while the remaining three (IDs\,7, 45, and 56) have values comprised between 99 and 310 cm$^{-3}$. 
Overall, there is no indication that the integrated line emission is dominated by regions of density higher that 100 cm$^{-3}$,  although some high-density clumps may still be present.\\

We compare the dust reddening obtained from the Balmer decrement with that computed using the Paschen lines in Fig. \ref{fig:attenuation}. The left panel shows that the E(B-V) values derived from the ratio of Paschen lines to \hb\ agree well, within the 1$-\sigma$ scatter along the 1:1 relation, with the reddening inferred from the Balmer decrement. We observe that a high S/N (indicated by larger symbols) ensures better agreement among the different measurements. For high S/N ratio, the observed Pa$9$/\hb\ and Pa$9$/\ha\ ratios, corrected by dust attenuation assuming the \cite{Odonnell1994} attenuation curve (dotted gray line), agree well with their theoretical case B hydrogen recombination values, whereas a larger discrepancy is observed for low S/N (central panel of Fig. \ref{fig:attenuation}). This is due to the increased noise in the redder part of the MUSE spectrum, which makes detecting the continuum more challenging and increases the uncertainty in line measurements. This is, in turn, reflected on the inferred E(B-V) values (right panel).\\

\section{Dust maps}

\begin{figure}[!h]
\centering
\includegraphics[width=0.5\textwidth, trim=20 100 0 80, clip]{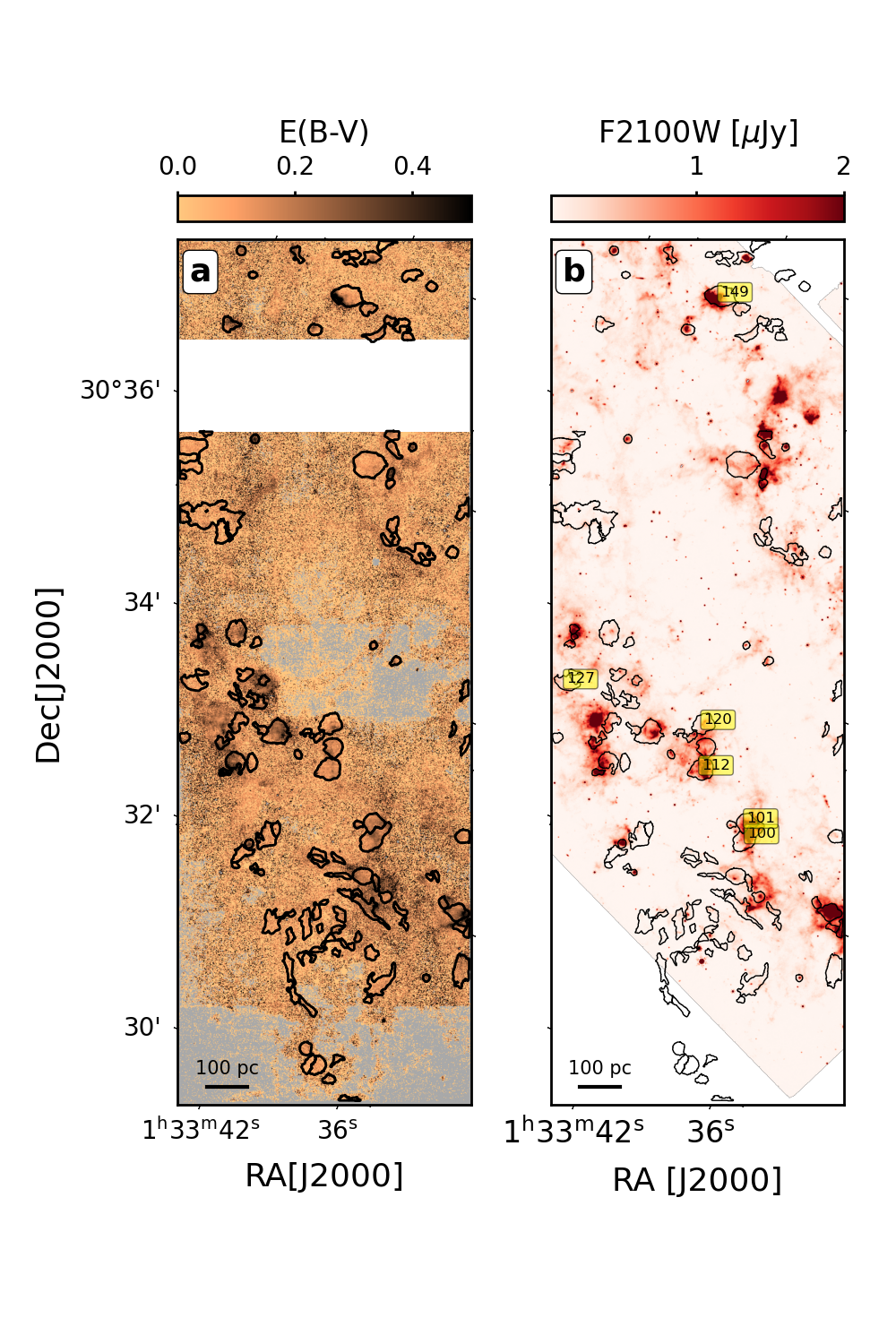}
  \caption{Two dimensional dust maps of M33: E(B-V) and MIRI F2100W maps in panel a and b, respectively. As in Fig. \ref{fig:line_ratio_maps} black contours represent the nebular masks classified as \hii\ regions. In panel b, numbered labels indicate specific nebular masks discussed in the main text.}
        \label{fig:dust_maps}
  \end{figure}

The E(B-V) map (Fig.\ref{fig:dust_maps}a) computed using the Balmer ratio, as described in Sect. \ref{sec:hii_int}, shows relatively moderate E(B-V) variations, with few heavily attenuated regions (E(B-V) > 0.3). Some of these structures are compact clumps within \hii\ regions, while others are more extended dusty structures distributed around several \hii\ regions. Fig. \ref{fig:dust_maps}b compares the location of the nebular masks identified in Sect. \ref{sec:astrodendro} and the mid-infrared 21$\mu$m image obtained with the JWST/MIRI instrument \citep{Wright2023} under the JWST GO program 2128 (PI: E. Rosolowsky).

\FloatBarrier
\section{Selected \hii\ regions}\label{app:AppG}

\begin{figure}
    \centering
    \begin{subfigure}{0.4\textwidth}
        \includegraphics[width=\textwidth,  trim=0 50 70 50, clip]{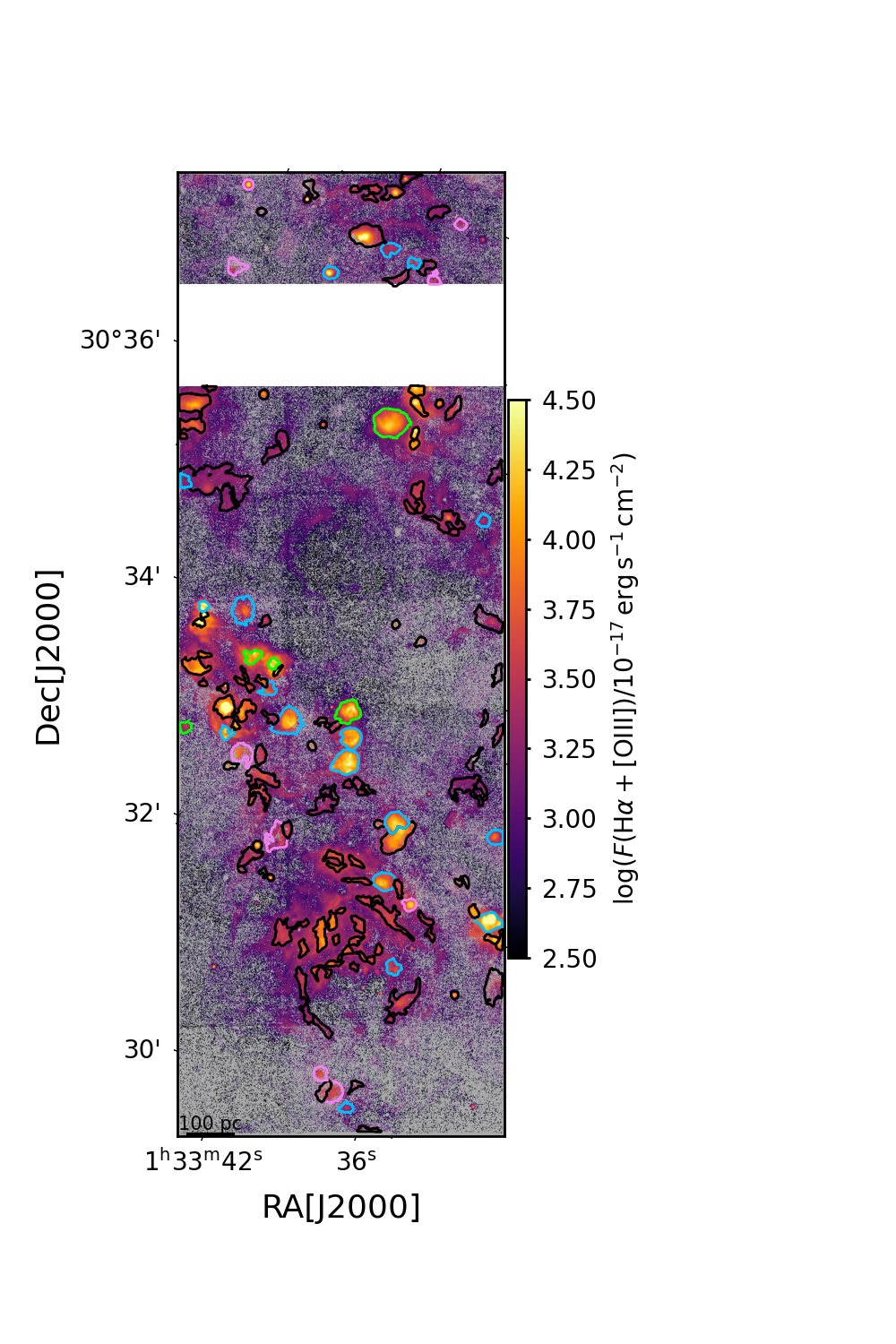}
    \end{subfigure}
    \begin{subfigure}{0.35\textwidth}
        \includegraphics[width=\textwidth]{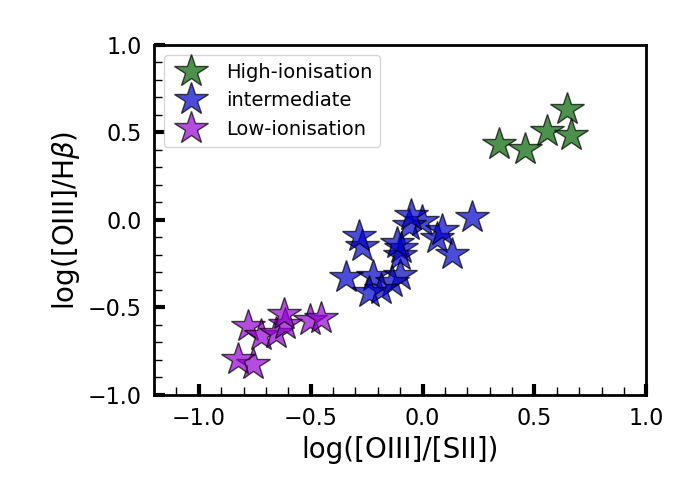}
    \end{subfigure}
    \caption{Spatial distribution of the 32 \hii\ regions selected to study the meadian radial profiles in Sect. \ref{sec:radial_tot} (top panel) and their position in the \oiii/\hb\ verus \oiii/\sii\ diagram (bottom panel). Violet, blue and green colours refer to the  `low-ionisation', `intermediate' and `high-ionisation' subsets, respectively.}
    \label{fig:AppG}
\end{figure}

We show the spatial distribution of the three subsets of the 32 \hii\ regions selected and described in Sect. \ref{sec:radial_tot} in Fig. \ref{fig:AppG} (top panel).
These subsets have been divided based on the \oiii/\hb\ and \oiii/\sii\ ratios in `low-ionisation', `intermediate', and `high-ionisation' subsets, as illustrated in the line ratio diagrams in Fig. \ref{fig:AppG} (bottom panel).

\end{appendix}

\end{document}